\documentclass[preprint2]{aastex}
\newcommand{\izw}{I\,Zw\,18}

\newcommand{\simgt}{\lower.5ex\hbox{$\;\buildrel>\over\sim\;$}}
\newcommand{\simlt}{\lower.5ex\hbox{$\;\buildrel<\over\sim\;$}}
\newcommand{\magsq}{mag~arcsec$^{-2}$}
\newcommand{\hst}{{\sl HST}}
\newcommand{\ew}{{\sl EW}}
\newcommand{\chimin}{{$\chi^2_{\rm min}$}}
\newcommand{\gchimin}{{$\Sigma\,\chi^2_{\rm min}$}}
\slugcomment{Submitted to ApJ}
\shorttitle{Stellar Populations in I\,Zw\,18}
\shortauthors{Hunt, Thuan, \& Izotov }
\received{2002 May 12}
\begin{document}

\title{New Light on the Stellar Populations in \izw: Deep Near-Infrared Imaging}

\author{Leslie K. Hunt}
\affil{I.R.A.-Firenze/C.N.R., Largo E. Fermi 5, I-50125 Firenze, Italy}
\email{hunt@arcetri.astro.it}

\author{Trinh X. Thuan}
\affil{Astronomy Department, University of Virginia, Charlottesville, VA 22903, USA}
\email{txt@virigina.edu}

\and

\author{Yuri I. Izotov}
\affil{Main Astronomical Observatory, Ukrainian National Academy of Sciences,
27 Zabolotnoho, Kyiv 03680, Ukraine}
\email{izotov@mao.kiev.ua}

\begin{abstract}
We present deep $JHK$ images of \izw, the most metal-deficient 
Blue Compact Dwarf (BCD) galaxy known, and analyze
them in conjunction with archival \hst/WFPC2 optical images.
To investigate the stellar populations, we have
divided the main body of \izw\ into eight regions, and fit the optical,
near-infrared (NIR), and hybrid optical-NIR colors of these and the C
component with evolutionary synthesis models. 
The composite best fit is obtained for an age for evolved stellar
populations of $\simlt$ 200\,Myr;
fits with an older age of $\simlt$ 500\,Myr are less likely but possible.
Our data show no evidence for stellar populations in \izw\ older than this,
although as much as 22\% of the stellar mass in older stars (4\% in $J$ light)
could remain undetected.
The colors of the young and intermediate-age
stellar populations are significantly affected by widespread and 
inhomogeneously distributed ionized gas and dust. Ionized gas emission is 
important in every region examined except the NW star cluster.
Extinction is significant in both the NW and SE clusters.
Red 
$H-K$, $B-H$, and $V-K$ colors are not reliable indicators of old stellar populations 
because ionized gas emission is also red in these colors.
$V-I$, on the other hand, reliably separates stars from gas because the
former are red ($V-I$ $\geq$ 0.4) while the latter is blue ($V-I$ $\sim$ --0.4).

\end{abstract}


\section{Introduction}

The question of whether there are young galaxies in the local 
universe forming stars for the first time  
is of considerable interest for galaxy formation and cosmological 
studies. There are several reasons for this.
First, Cold Dark Matter models predict that low-mass dwarf galaxies 
could still be forming at the present epoch because they originate 
from density fluctuations considerably smaller than those 
giving rise to the giant ones. Thus the existence of young dwarf galaxies 
in the local universe would put strong constraints on the 
primordial density fluctuation spectrum. 
Second, while much progress has been made in 
finding large populations of galaxies at high ($z$ $\geq$ 3) redshifts 
\citep{steidel,dey,papovich}, truly young galaxies in the 
process of forming remain elusive in the distant universe. 
The spectra of those far-away galaxies 
generally indicate the presence of a substantial amount of heavy elements, 
implying previous star formation and metal enrichment. Thus it is important
to have examples of truly young galaxies in the local universe because 
they can be used as laboratories to study star formation and 
chemical enrichment processes in 
environments that are sometimes much more pristine than those in known 
high-redshift galaxies. Moreover, their proximity allows studies of their 
structure, metal content, and stellar populations with a sensitivity, 
precision, and spatial resolution that faint distant high-redshift galaxies 
do not allow. Finally, in the hierarchical model of galaxy formation,
large galaxies result from the merging of smaller structures. 
These building-block galaxies are too faint and small to be studied at 
high redshifts, while we stand a much better chance of understanding them 
if we can find some local examples.

The Blue Compact Dwarf (BCD) galaxy \izw\  is one of the best candidates 
for being a truly young galaxy. It is 
described by \citet{zwicky} as a double system of compact galaxies,
which are in fact two bright knots of star formation separated by 
5\arcsec\ and referred to as the brighter northwest (NW) 
and fainter southeast (SE) components; these two components comprise the main body.
Later studies revealed a blue irregular star-forming region 
$\sim$ 22\arcsec\ northwest of the NW 
component, referred to as component C. \citet{dufoura},
\citet{vanzee} and \citet{it98} have shown the C component
to have a systemic radial velocity equal to that of the ionized gas in the 
NW and SE components, thus establishing its physical association with \izw.
Furthermore, 21 cm VLA mapping by \citet{vanzee} has shown that this 
component is embedded within a common H I envelope with the main body.

As the lowest-metallicity galaxy known, \izw\  has been studied
intensively over the thirty years since 
the first determination of its heavy element abundance \citep{ss70}.
The low oxygen abundance of 1/50 the solar value of 
\izw\  has been confirmed by 
many groups \citep [~and references therein]{izotov01}. 
This lack of chemical enrichment has led to the idea that \izw\
may be a young galaxy, only now forming stars for the first time
\citep{ss72}.  
\citet{it99} have argued that the  
heavy element abundance of a BCD is a good age indicator. They 
found that all BCDs with heavy 
element mass fraction $Z$ $\leq$ $Z_\odot$/20  
show constant C/O and N/O abundance ratios, and interpret this constancy as 
meaning that C and N in these most metal-deficient galaxies are made 
in the same massive stars that produce O, and that intermediate-mass stars 
(3 $M_\odot$ $\leq$ $M$ $<$ 9 $M_\odot$) have not had time to die and 
release their C and N production. This puts an age upper limit of the order 
of 400 Myr for these extremely metal-deficient BCDs \citep{it02}.
It was first thought that \izw\  did not follow that trend as 
\citet{garnett} have derived a C/O ratio for \izw\  that is 
considerably higher than in other metal-poor BCDs, and concluded that  
carbon in \izw\ has been enriched by an older generation of stars. 
However, \citet{it99} have 
rederived the C/O abundance ratio in \izw\  and found a value in 
agreement with that of other BCDs with $Z$ $\le$ $Z_\odot$/20.  
Therefore, no preceding low-mass carbon-producing stellar population
needs to be invoked. 
Young dynamical ages have been inferred for \izw\ from
the structure of expanding superbubbles of ionized gas driven by 
supernova explosions (15~--~27\,Myr: \citep{martin}; 13~--~15\,Myr: \citep{dufour}).  

The conclusion that \izw\  might be a young primeval galaxy in the local 
universe, with age $\leq$ 400 Myr, has been challenged, however,  
by the authors of several recent color-magnitude 
diagram (CMD) studies of stars resolved in \hst\ images of \izw. 
In the first CMD study based on $U$, $V$, and $I$ \hst/WFPC2 images,
\citet{ht95} concluded that the optical colors of the stars in \izw\ 
were consistent with those of young massive main sequence 
stars, and the colors of the unresolved emission with an 
intermediate-age stellar population of B and early A stars. But 
the data did not go deep enough to detect individual evolved stars. 

From an independent
 set of $B$, $V$ and $R$ \hst/WFPC2 images,
\cite{dufour} concluded that the current star formation 
in the main body began at least 
30 -- 50 Myr ago, continuing to the present, consistent 
with the dynamical ages derived from expanding superbubbles of ionized gas. 
As for the C component, \cite{dufour} found that it
consists of an older stellar population composed of faint red stars
 with ages 100 -- 300 Myr on 
which the current modest starburst (blue stars with age $\sim$ 40 Myr) is  
superimposed in its southeastern half. 
Those ages are consistent with those derived from 
heavy element abundance ratios \citep{it99}. 

However, 
\citet{aloisi} have reanalyzed the two \hst/WFPC2 data sets 
above and, going deeper thanks to improved reduction techniques, 
concluded that there were two episodes of star formation in the main body:
a first episode occurring over the last 0.5 -- 1 Gyr, an age more than 10 
times larger than that derived by \citet{dufour}, and a second episode 
with more intense activity taking place between 15 and 20 Myr ago, with  
no star formation occurring within the last 15 Myr. For the C component, 
\citet{aloisi} estimated an age not exceeding 200~Myr. 
Subsequently, \citet{ostlin} has carried out a near-infrared (NIR) 
CMD study based on \hst/NICMOS $J$ ({\sl F110W}) 
and $H$ ({\sl F160W}) images. He concluded that 
the main body of \izw\ is dominated by 
two populations, one 10--20\,Myr population of red supergiants and another
considerably older  
0.1--5\,Gyr population of asymptotic giant branch (AGB) 
and red-giant stars. Thus, it appears that CMDs require two episodes 
of star formation in \izw: an old 
burst, responsible for the red faint stars, and a more recent one which
produces the red supergiant and blue massive stars seen to dominate the CMDs.
More recently, \citet{recchi} also found that two bursts were necessary
to explain the dynamical and chemical evolution in \izw: an older burst which occurred
$\sim\,$300\,Myr ago, and a more recent one with age between 4--7\,Myr.

\citet{izotov01} and \citet{ko01} have pointed out that 
comparison of observed colors with those predicted by burst models of star
formation inevitably leads to an underestimate of the age,
and an overestimate of the star-formation rate.
It is therefore important to consider various star formation scenarios 
(such as continuous star formation) in any comparison between models and 
observations.
\citet{legrand00} and \citet{legrand00p} have gone to the other 
extreme by arguing that
star formation in \izw\ may proceed, not in episodic bursts, but in 
a continuous fashion over a long period of many Gyr, 
at the low rate of about $10^{-4} M_\odot$\,yr$^{-1}$,
more than a 1000 times lower than that inferred by \citet{aloisi}.
This continuous low star formation would  
result in an extended underlying low surface brightness stellar 
component in \izw. However, \citet{izotov01} 
and \cite{papa02} have shown that the 
existence of such a component is supported neither by spectroscopic nor 
photometric measurements of the extended emission around the star forming 
regions in \izw. 

It is thus clear that, in spite of numerous studies,
there is still no consensus regarding the age of \izw\  and 
its star formation history. 
While CMD analysis is the method of choice for the determination of
stellar ages, it is also subject to several uncertainties  
such as distance determination,
extinction, and contamination of the stellar colors by gaseous emission. 
The NIR bands provide a clear advantage over optical bands in this
context, because of their enhanced sensitivity to evolved red stars 
and reduced effects of dust extinction.
To revisit the question of the ages of the stellar populations in
\izw, we have acquired deep $JHK$ images with UKIRT/IRCAM3.
Using these together with \hst\ WFPC2 and NICMOS archival images, 
we can directly 
address the above questions of extinction and 
gas contamination. Moreover, we can derive the color 
transformations between the \hst\ and ground-based filters. 
This is necessary because the \hst\ NICMOS
1\micron\ filter ({\sl F110W}) is very different from the standard 
Johnson $J$ filter.
By necessity, we shall focus here on average stellar 
surface brightness distributions
rather than on luminosities of single stars.
While seemingly a liability, such an approach will allow 
us a more complete examination of 
the stellar content of the galaxy and its age.

The observations and analysis techniques are presented in Section \ref{obs}.
The derived NIR and hybrid optical-NIR colors are 
discussed in Section \ref{broadband}.
In Section \ref{age}, we model the broadband colors with a variety of
star-formation histories, and discuss the results in terms of
stellar population ages, gaseous emission and dust extinction. 
We summarize our conclusions in Section \ref{concl}. 

\section{Observations and Analysis\label{obs}}

We acquired $J$ (1.2\micron), $H$ (1.6\micron), and $K$ (2.2\micron)
images of \izw, including the C component, with the 3.8-m United Kingdom Infrared Telescope
(UKIRT\footnote{The United Kingdom Infrared Telescope is operated by the
Joint Astronomy Centre on behalf of the U.K. Particle Physics
and Astronomy Research Council.})
equipped with IRCAM3.
The observations were performed in the UKIRT Service Observing program
in March, 1999,
as part of our ongoing project of NIR imaging and spectroscopy of BCDs.
The IRCAM3 plate scale is 0\farcs28 per pixel, 
with a total field-of-view of 72\arcsec~$\times$~72\arcsec.
To obtain a clean image of the C component, 22\arcsec\ to the NW,
source and empty sky positions were alternated,
beginning and ending each observing sequence with a sky position.
Before the beginning of each sequence,
dark exposures were acquired with the same parameters as the subsequent
science frames.
Total on-source integration times
were 43\,min in $K$, 26\,min in $H$, and 39\,min in $J$.
Individual frames were dark-subtracted and flat-fielded with 
the average of adjacent empty sky frames,
after editing them for stars (to avoid ``holes'' in the reduced frames)
and applying an average-sigma clipping algorithm.
The reduced frames were then aligned and averaged.
All data reduction was carried out in the 
IRAF environment\footnote{IRAF is the Image Analysis and Reduction Facility
made available to the astronomical community by the National Optical
Astronomy Observatory, which is operated by AURA, Inc., under
contract with the U.S. National Science Foundation.}.

Photometric calibration was performed by observing standard stars
from the UKIRT Faint Standard List \citep{hawarden}
before and after the source observations.
Each standard star was measured in several different positions on the array,
and flat-fielded by dividing the clipped mean of the remaining frames 
in the sequence. To correct the standard-star 
photometry for atmospheric extinction,
we used the UKIRT mean extinction coefficients of 
0.102, 0.059, and 0.088 mag/airmass for $J$, $H$, and $K$ respectively.
Formal photometric accuracy, as measured by the dispersion in 
the standard star magnitudes is 0.025\,mag in $J$ and $H$ and 
0.04\,mag in $K$.

\begin{figure}
\plottwo{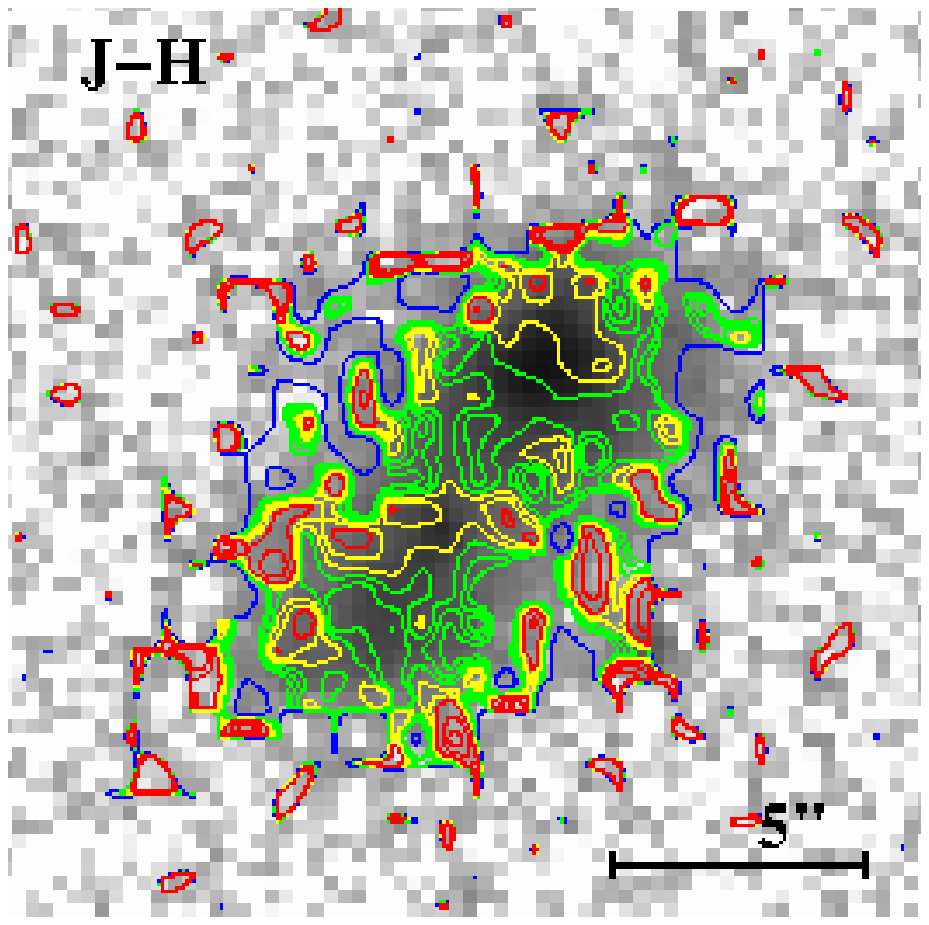}{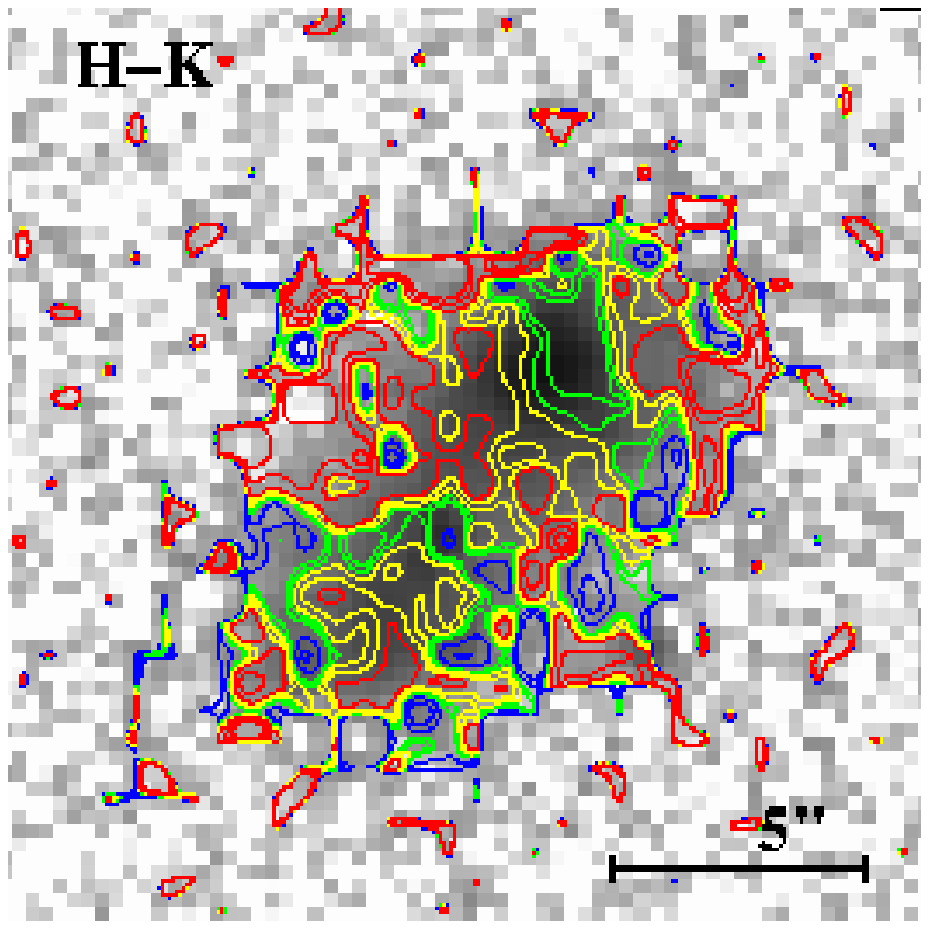}
\caption{$J$ image shown in grey scale, with NIR color contours
superimposed (North is up, East to the left).
The grey scale runs from 18 to 25 $J$ \magsq, with the brightest regions
appearing darkest.
In the left panel, $J-H$ is contoured as follows:
$J-H$~=~--0.5 to 0.0 ({\it black [blue]} contours); 
$J-H$~=~0.0 to 0.3 ({\it light grey [green]} contours); 
$J-H$~=~0.3 to 0.5 ({\it white [yellow]} contours); and 
$J-H$~=~0.5 to 1.2 ({\it dark grey [red]} contours).
In the right panel, 
$H-K$~=~--0.5 to 0.0 ({\it black [blue]} contours); 
$H-K$~=~0.0 to 0.2 ({\it light grey [green]} contours); 
$H-K$~=~0.2 to 0.5 ({\it white [yellow]} contours); and 
$H-K$~=~0.5 to 1.2 ({\it dark grey [red]} contours).
The scale is shown in the lower right corner.
\label{fig:nir}}
\end{figure}

Color images were derived by registering the images to a few tenths of a 
pixel with a
cross-correlation algorithm, then subtracting the magnitude images.
The $J$ image is shown in Fig. \ref{fig:nir}, together
with $J-H$ (left panel) and $H-K$ (right panel) contours superimposed. 
These contours will be discussed in Section \ref{nircolors}.

\placefigure{fig:nir}

\subsection{\hst\ WFPC2 and NICMOS Images\label{hst}}

We retrieved from the \hst\ archives WFPC2 images of \izw\  in the
F555W ($V$) and F814W ($I$) filters (PI: D.A. Hunter, GO-5309),
and in the F439W filter ($B$) (PI: R. Dufour, GO-5434).
The pipeline reduction was used, followed by combination of the
individual images with the IRAF/STSDAS\footnote{STSDAS is the Space Telescope 
Science Data Analysis System.} task {\it crreject}.
The images were calibrated and transformed to ground-based $BVI$
according to the precepts of \citet{holtzmann}, then rotated to canonical
orientation (N up, E left).
The $V-I$ image was obtained by subtracting the magnitude images.

\placefigure{fig:opt}

\begin{figure}
\plottwo{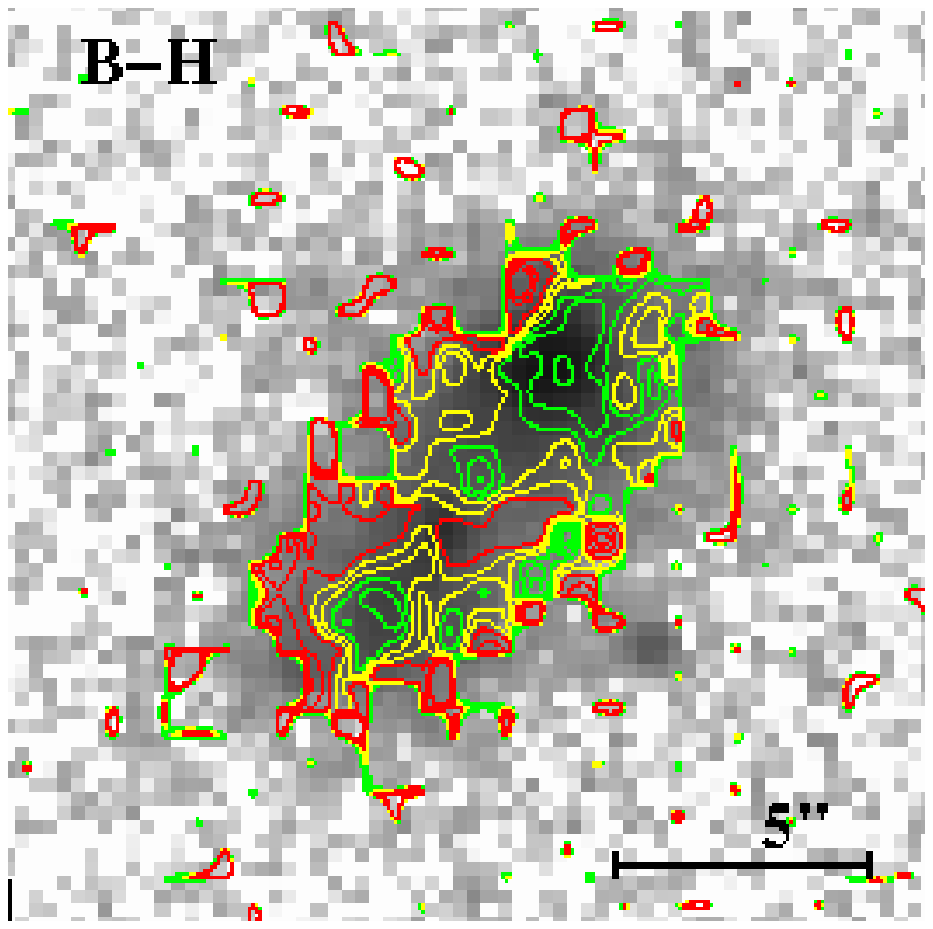}{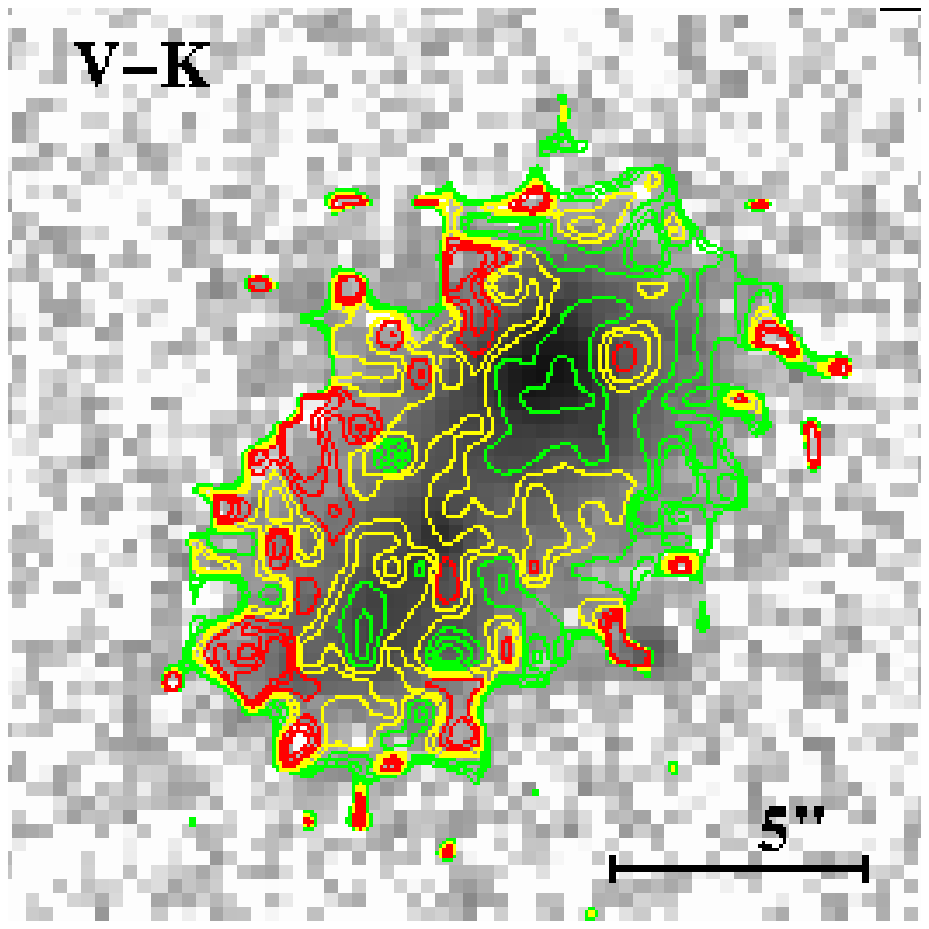}
\caption{Hybrid optical-NIR color images of \izw:
$B-H$ is shown in the left panel, and $V-K$ in the right.
$B-H$ is contoured as follows: $B-H~<~0.7$ ({\it light grey [green]} contours); 
$B-H$~=~$0.7$ to 1.2 ({\it white [yellow]} contours);
and $B-H$~=~1.2 to 2.0 ({\it dark grey [red]} contours).
$V-K$ is contoured as follows:
$V-K~=~-0.1$ to 1.0 ({\it light grey [green]} contours);
$V-K$~=~$1.0$ to 1.5 ({\it white [yellow]} contours);
and $V-K$~=~1.5 to 2.0. ({\it dark grey [red]} contours).
North is up, East to the left, and the spatial scale is shown in the lower
right corner of each panel.
\label{fig:hybrid}}
\end{figure}

Hybrid ground-based NIR--optical WFPC2 color images were derived by first 
rotating them to the usual orientation (North up and East left),
rebinning them to the UKIRT pixel size, and finally aligning them 
to a fraction of a pixel by cross-correlation.
These color images, shown as contours superimposed on the $J$ image, are shown
in Fig. \ref{fig:hybrid}.

\placefigure{fig:hybrid}

The NICMOS {\sl F110W}\ and {\sl F160W}\ images were also
acquired from the \hst\ archive (PI: G. \"Ostlin, GO-7461), and 
re-reduced with STSDAS task {\it calnica} using the best reference files.
They were corrected for the pedestal effect with the software
of van der 
Marel\footnote{\it http://www.stsci.edu/$_{\tilde{\,}}$marel/software/pedestal.html}, 
and calibrated in the standard way.
Finally, they were rotated to canonical orientation.
The analytical color transformation of \citet{origlia}
was then applied to the rotated {\sl F110W}--{\sl F160W}\ color image to transform it
to the standard $J-H$.
To compare the transformed \hst/NICMOS image with our 
ground-based $J-H$ image, we rebinned the \hst\ image to  
the larger UKIRT pixel size, and registered the two by cross-correlation.

\subsection{Aperture Photometry \label{phot}}

Aperture photometry with varying aperture diameters was performed 
on the NIR UKIRT and optical WFPC2 images 
at the locations of the two main star-forming regions in the main body of \izw,
the NW and the SE brightness peaks.
To better study color gradients and possible changes in gaseous content 
and stellar populations over short distances in \izw, 
we also performed 
photometry in circular apertures of radius 1\arcsec\  along a position
angle (PA) of 149$^\circ$, E from N, aligned with the NW and SE brightness peaks. 
Besides the NW and SE peaks, photometry was also obtained for
the intermediate region between the NW and SE regions (``Inter''), 
two extended regions to the NW of the NW peak (``NW1'' and ``NW2''), 
and the two extended regions to the SE of the SE peak (``SE1'' and ``SE2'').
We used an aperture of radius 1\farcs5 in the NW2 and SE2 regions because of lower
signal-to-noise.
The different apertures and their designations are shown 
in Fig. \ref{fig:opt} superimposed on the \hst\ $I$ image in the left
panel, and on the $V-I$ image in the right; the photometry itself, together
with larger-aperture photometry for all regions, is reported in 
Table \ref{table:phot}.

We have also obtained a UKIRT NIR image of the C component,
and performed aperture photometry
on two positions: one centered on the main brightness peak denoted by 
``C (central)'', and the other on the extended emission toward the NW denoted 
``C (extended)''.
This photometry, together with the global photometry for the C component,
is also reported in Table \ref{table:phot}.

\subsection{Surface-Brightness Profiles and Cuts\label{cuts}}

We have fit ellipses to the $J$ image of \izw, with
the centers defined by the center of symmetry of the outer regions, 
which lies roughly halfway between the NW and SE emission peaks.
The PA and ellipticity of the ellipses were fixed
to 141$^\circ$ (E from N) and 0.23, respectively, which were the values 
the isophotal fitting converged to.
Note that this PA is 8$^\circ$ smaller than that we used for the cuts;
the latter was defined by eye to contain the maxima of the NW and SE brightness
peaks.
Using this geometry, $J, H, K$, $B$, $V$, and $I$ radial profiles were derived,
and are shown in Fig. \ref{fig:pro}.
Inspection of the figure shows that our $K$ image, when elliptically
averaged, achieves a limiting surface brightness of $\sim$\,24 \magsq, 
enabling us to search effectively for an extended underlying old stellar 
population.

The mean NIR colors of extended emission were derived by calculating
the average of all points in the profile with $\mu(K)\,<\,23$\,\magsq,
$\sigma_K\,\leq\,0.3$, and radius $R\,>$\,3\farcs5.
The mean optical and hybrid colors were derived in a similar way, with
$\mu(V)\,\leq\,27$\,\magsq\ and radius $R\,>$\,5\farcs5.
The mean colors of the extended emission turn out to be:
$J-H\,=\,0.38\,\pm\,0.13$, $H-K\,=\,0.32\,\pm\,0.19$,
$V-I\,=\,-0.43\,\pm\,0.25$,
$B-H\,=\,1.77\,\pm\,0.37$,
$V-K\,=\,1.69\,\pm\,0.38$,
where the quoted uncertainties are 
the standard deviations of the ensemble, rather than 
photometric ``errors''.
These values have been corrected for Galactic extinction according
to the \citet{schlegel} value $A_B$\,=\,0.138, and with the extinction
law of \citet{cardelli}.
The mean colors of the extended emission are indicated 
in Fig. \ref{fig:pro} by horizontal dashed lines. 
The peak at $R$\,$<$\,2\arcsec\ corresponds to the approximate
superposition of the NW and SE peaks in the elliptical average.

\placefigure{fig:pro}

Although elliptical averaging is important because of the gains that
can be achieved in signal-to-noise ratio, the inner structure of
\izw\ is clearly not symmetric. 
Therefore, to better quantitatively examine color gradients, 
we have also derived surface-brightness profiles along 
a ``cut'', coincident with the orientation of the aperture photometry
(PA\,=\,149$^\circ$ and aligned with the two main brightness peaks).
The width of the cut is 1\farcs7 in the perpendicular direction
(6 pixels for the NIR images, 38 pixels for WFPC2).
The resulting brightness and color profiles are shown in Fig. \ref{fig:cut}, where
positive radii are towards the NW, negative towards the SE;
the origin is defined by the NW brightness peak.
NIR surface-brightness and color profiles are shown in the left panels, 
and analogous hybrid NIR-optical profiles in the right panels.
Cut regions are labelled with their regional designation
(see $\S$\ref{phot}) at their approximate radial distance.
The colors in the 2\arcsec\ 
apertures are indicated by short horizontal dashed (red) lines. 
The color of the extended region is indicated by the long
horizontal dashed (green) line\footnote{Although all figures are in 
black and white in the printed version
of the paper, some will appear in color in the on-line version. Here,
and for subsequent figures, we indicate in parentheses the colors.}.
All colors have been corrected for Galactic extinction as described
above.

\placefigure{fig:cut}

\section{Broadband Colors \label{broadband}}

By combining the optical \hst\ archival images together with our 
NIR data, we have at our disposal six broadband filters, $B$, $V$, $I$, $J$, $H$ 
and $K$, and thus five independent colors. 
We have chosen to analyze $V-I$, $J-H$, $H-K$, $B-H$, and $V-K$ for the following
reasons.
The optical $V-I$ color provides important information about gas
contamination, extinction, and young stars.
Moreover, it distinguishes very well ionized gas emission from evolved stars.
The NIR colors put constraints on 
the presence of an evolved stellar population, as well as on 
the presence of hot dust and ionized gas ($H-K$).
Hybrid optical-NIR color combinations, such as $B-H$ and $V-K$, are
effective extinction diagnostics: 
according to the \citet{cardelli} extinction curve
and \citet{holtzmann},
the reddenings for $V-K$ and $B-H$ are respectively $0.94\,A_V$ and 
$1.17\,A_V$,
large enough to allow the separation of the
effects of stellar population age gradients from those of dust.
These hybrid colors also provide important diagnostics
for the nature of the stellar populations. In particular, the 
$V-K$ color is very sensitive to age, as the onset of the AGB phase at $\sim$ 
100~Myr causes a sudden intense
reddening, making $V-K$ larger than $\sim$ 2 \citep{girardi}.
Hybrid optical-NIR colors are ambiguous however for discriminating 
stellar population effects from ionized gas contamination and
extinction effects since they are red in all cases.
To distinguish gas from older stars, the color of choice is $V-I$ and, to
a lesser extent $J-H$: while these colors are blue for gas, they are red
for older stars. 
We discuss in turn the optical, NIR, and hybrid colors and color
gradients in the following subsections.

\subsection{Optical Colors \label{optcolors}}

\begin{figure}
\plottwo{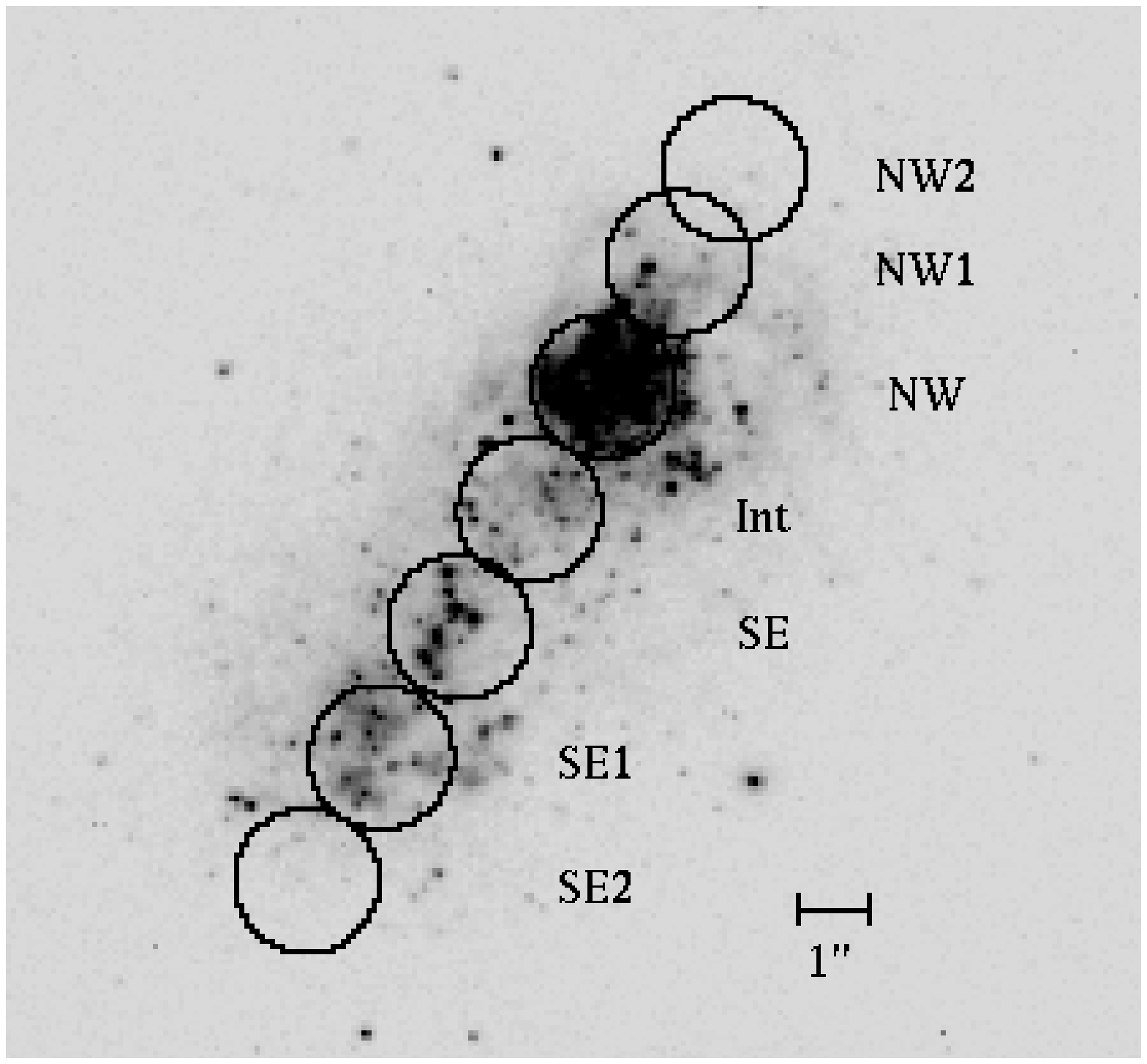}{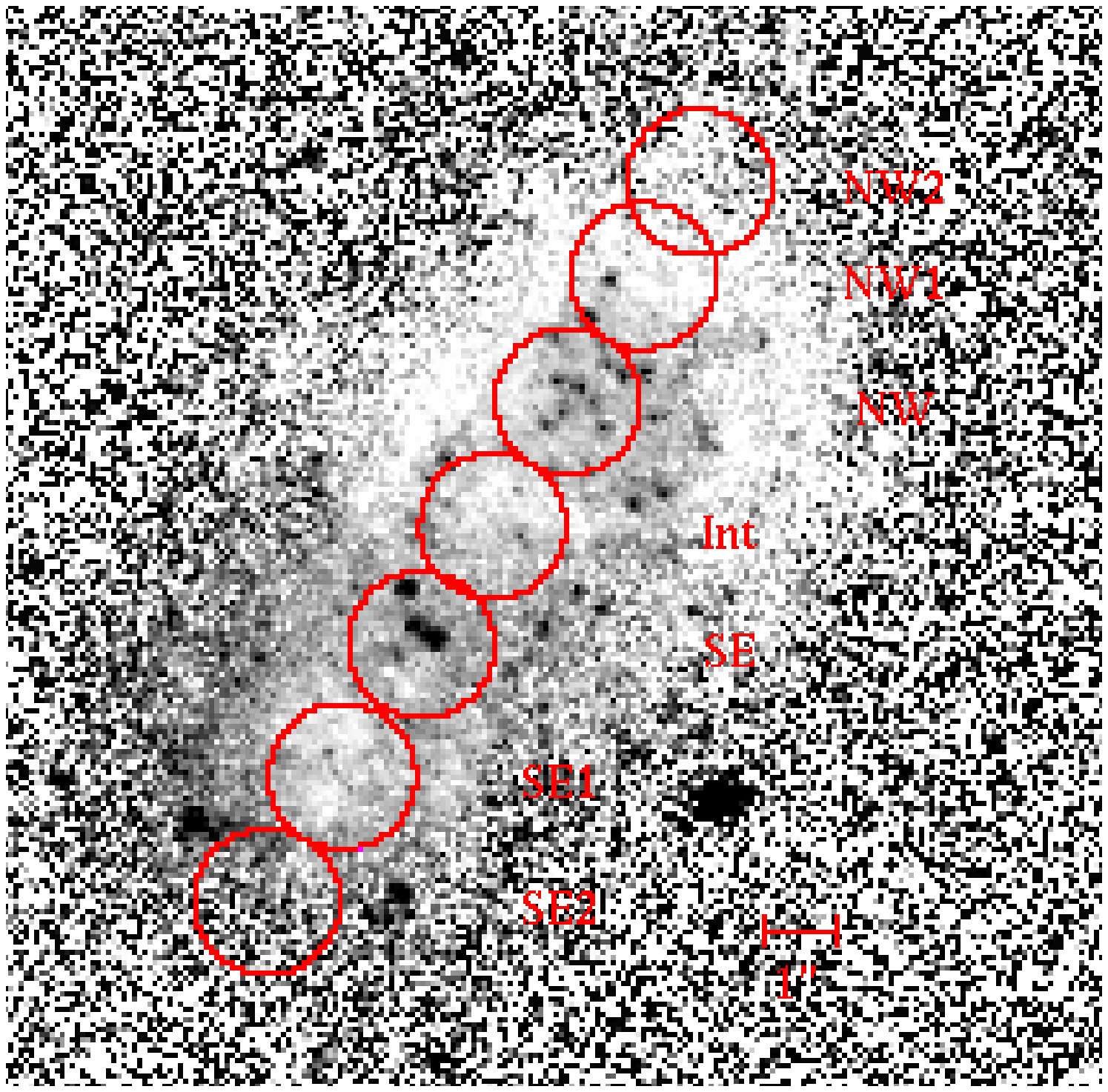}
\caption{The left panel shows the aperture positions for the 
various regions superimposed on the {\it F814W} image.
The right panel shows the $V-I$ color image for comparison;
blue colors are shown by white pixels, and red colors by black ones.
In both panels, North is up, East to the left, and
the spatial scale is shown in the lower right corner.
In the Figure, ``Int.'' denotes the Inter.(mediate) region as defined
in the text.
\label{fig:opt}}
\end{figure}

\begin{figure}
\plottwo{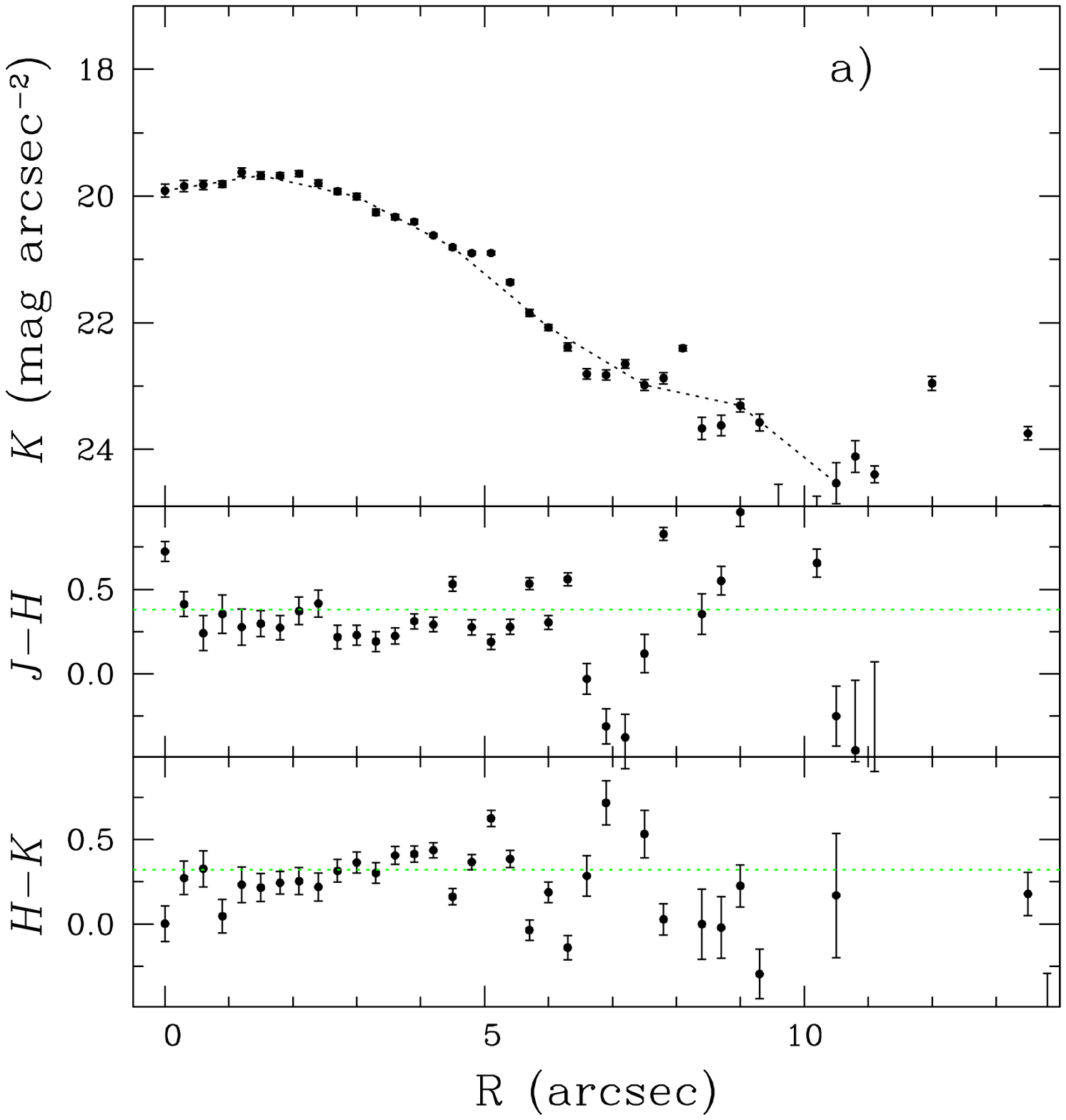}{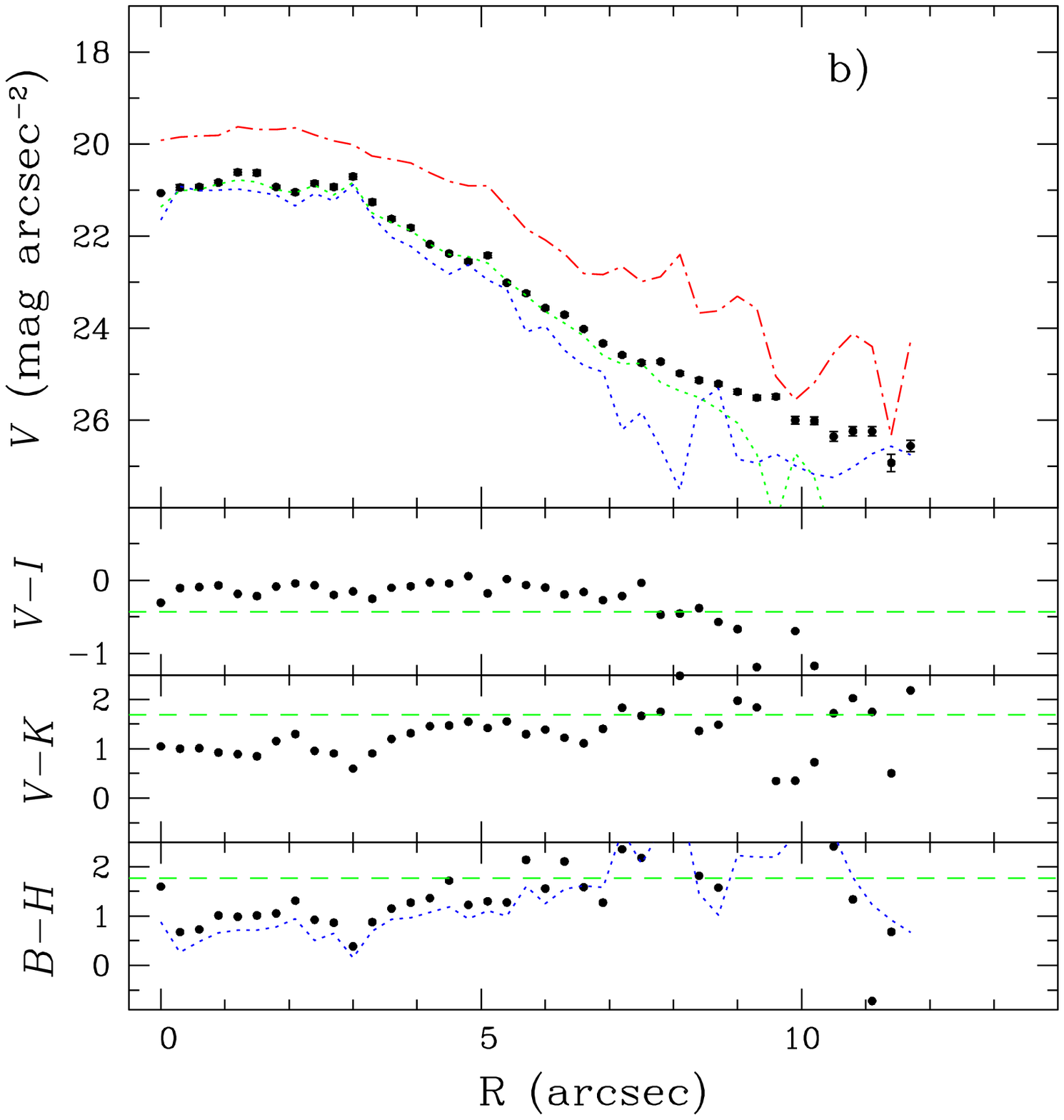}
\caption{Left panel a):
Elliptically-averaged surface-brightness profiles centered between
the NW and SE peaks.
Ellipticity is fixed to 0.23 and position angle to 141$^\circ$.
Only those points with $\sigma_K\,\leq\,$0.5 mag are plotted,
and the error bars show the root-mean-square deviation over the ellipse.
The left panel (a) shows the NIR profiles, and
the right panel (b) the optical/hybrid profiles.
The right panel 
The dashed horizontal lines in the lower panels are the mean colors
of the lower-surface brightness extended region as described in the text.
In the upper right panel ($V$), the $K$-band profile is shown as the uppermost
dot-dashed line, the $B$- and $I$-band profiles as the dotted lines nearest the data
($V$) points.
The $B$ profile starts to deviate from the other two optical colors at $R\sim$7\arcsec,
probably because it is a slightly less sensitive image.
Also on the right, the $B-J$ color is shown as a dotted curve in the
lowest ($B-H$) panel.
\label{fig:pro}}
\end{figure}

\begin{figure}
\plottwo{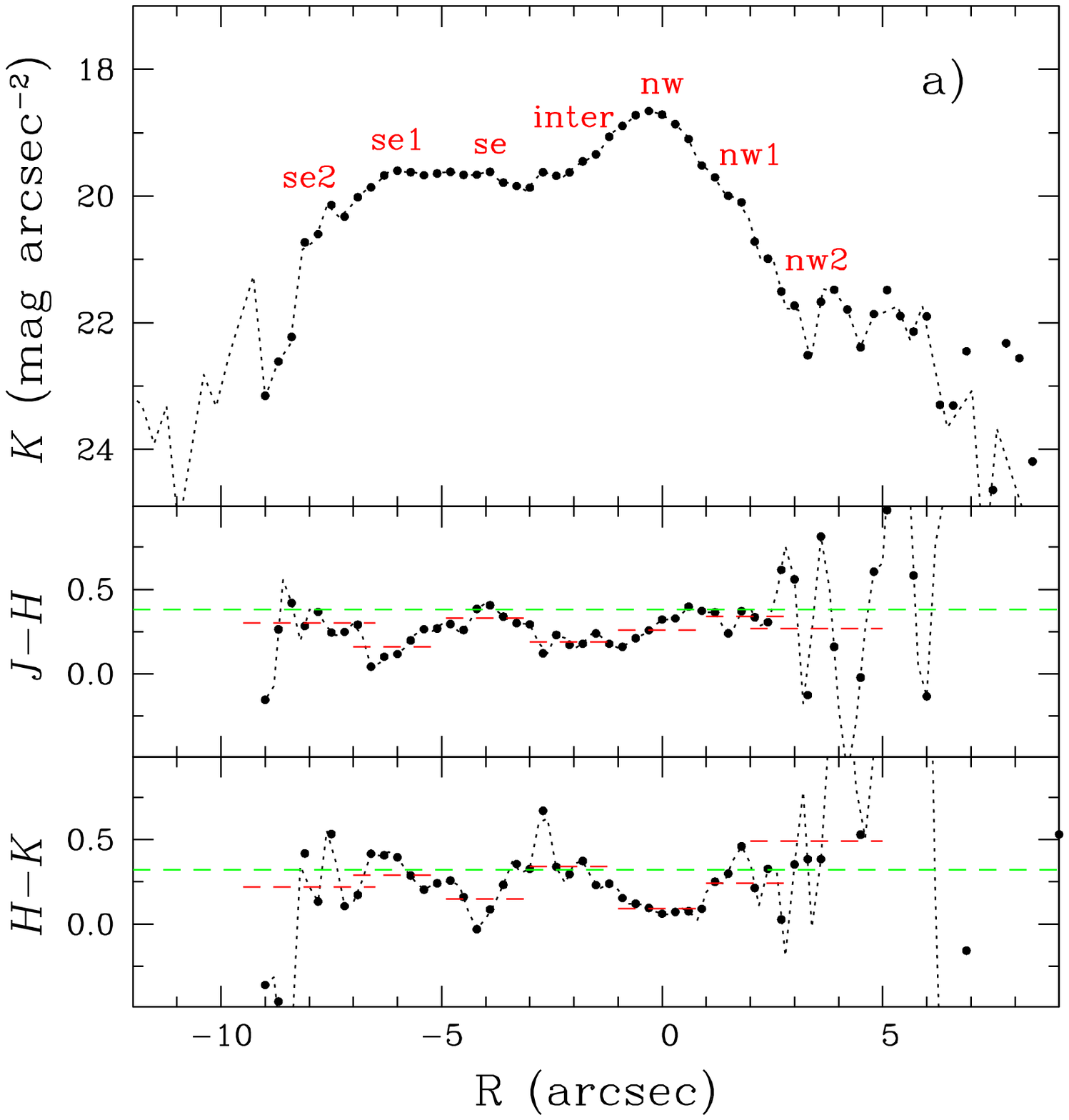}{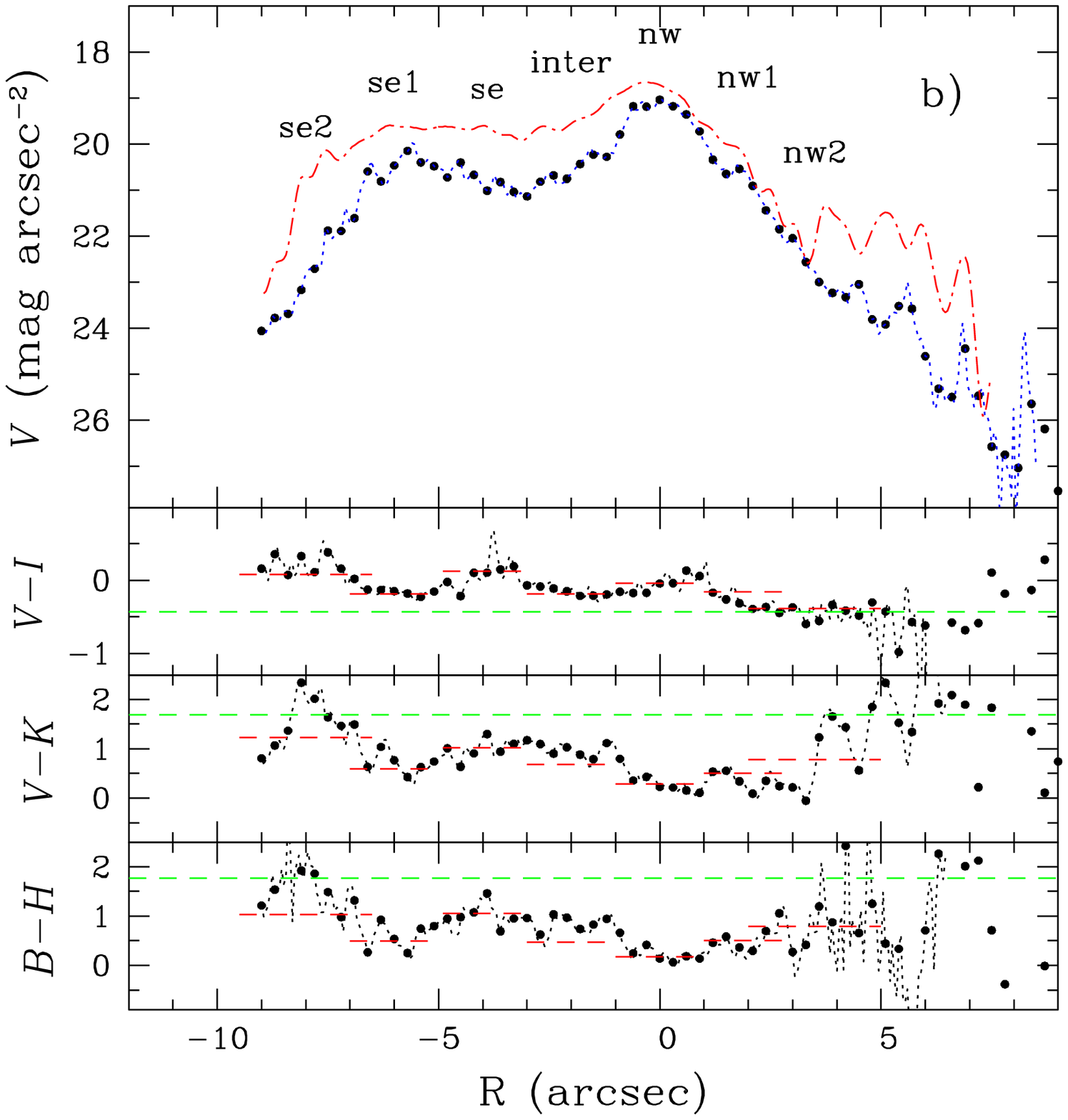}
\caption{Surface brightness ``cuts'' oriented at a position angle
of 149$^\circ$ along the NW and SE emission peaks.
The left panel (a) shows the NIR colors; 
the right panel (b) the optical/hybrid colors. 
Regions as described in the text are labelled. 
The $K$-band profile is reproduced (above the $V$-band profile) in
the right panel.
The short dashed horizontal lines in the lower panels are the colors
obtained for the various regions as described in the text; the long horizontal
dashed line is the mean color of the extended region.
\label{fig:cut}}
\end{figure}

Inspection of the $V-I$ color image (right panel of Fig. \ref{fig:opt}) shows that
there is a blue ``crown'' surrounding the NW star cluster,
a red border to the E of the SE cluster, and numerous
very red point sources scattered throughout the NW and SE brightness
peaks and the region separating them.
There is also a blue region around the SE1 star cluster. 
The $V-I$ colors in the crown range from
$-0.2$ to $-0.6$, and with redder colors toward the SE. 
The star clusters themselves appear to be relatively blue
($V-I\,<\,0$).
We ascribe the colors in the crown to ionized gas for the following three reasons.
First, the $V-I$ color of ionized gas is 
$\sim$ --0.4 at the redshift of \izw, similar to the crown color.
Second, in the \hst\ $I$ image (left panel of Fig. \ref{fig:opt}), 
filamentary structures can be seen surrounding the NW brightness peak, 
coincident with the blue crown. Third, examination of the H$\alpha$ maps 
of \citet{ostlin96} and \citet{izotov01} show very high equivalent widths at 
the location of the crown. 
This interpretation will be further discussed in the context of our color fits
in $\S$\ref{models}.

The extended emission, at $R\,\sim\,8$\arcsec\ and $\mu(V)\,=\,24.5 $ \magsq,
shows also very blue $V-I\,\sim\,-0.4$.
Such blue $V-I$ colors are highly unusual in BCDs. They 
are bluer than in any of the
H {\sc ii} galaxies studied by \citet{telles} but are comparable to those 
in SBS\,0335--052
\citep{thuan97}, another extremely metal-poor BCD 
(1/40 solar metallicity).

A striking feature of the $V-I$ map shown in the right panel of Fig. \ref{fig:opt}
is the patchiness of the
color, which can be most probably attributed to spatial variations in the 
dust and ionized gas emission.
In addition to the patchiness on small scales, there are also  
large-scale color gradients. These can be 
clearly seen in the cuts shown in the right panel of 
Fig. \ref{fig:cut}. There is a general reddening of $V-I$
going from the NW to the SE:
$V-I$ is $\sim$ --0.6 at the northernmost tip 
of the extended region to the NW,
and gets progressively redder going toward the SE where, outside the
main brightness peak, $V-I$ is $\sim$ 0.2.
In addition to this general trend, there are local red maxima in $V-I$
near the NW and SE clusters, and further to the S (SE2).
The red features evident in the $V-I$ color image (right panel of Figure \ref{fig:opt}) 
toward the SE can be seen in Fig. \ref{fig:cut}
at a radial distance of roughly $-8$\arcsec, and the
blue $V-I$ of the filaments toward the NW at a radial distance of 
$\sim\,5$\arcsec.

The high resolution of the optical color images also reveals numerous
very red point sources, particularly in the NW star cluster, where 
$V-I$ can be as high as 1.5 locally.
Even higher $V-I\,\sim\,1.7$ is seen in the reddest compact sources between 
the two brightness peaks.
Such red $V-I$ colors are consistent with those of bright extremely late-type stars
($>$\,K5) \citep{bb88}, presumably red supergiants, or with those of dusty H {\sc ii} 
regions.

\subsection{NIR Colors \label{nircolors}}

The $J-H$ color image shown in the left panel of Fig. \ref{fig:nir}
also reveals color gradients, although most of the image
has $J-H\,\sim\,0.2$ (light grey\footnote{In the online color version,
these and following contours correspond to: 
black=blue, light grey=green, white=yellow, dark grey=red.} contours), 
the colors of young or intermediate-age stars, assuming 
there is no contamination from ionized gas emission.
Moderately red $J-H$ ($\sim$ 0.4, white contours) is seen towards 
the N of the NW cluster, and along the N border of the SE cluster.
Very red $J-H$ structures ($J-H\,\sim\, 1.0$, dark grey
contours) lie toward the E and the SE, both outside of the SE brightness peak.
Toward the E, $V-I\,<\,0$ and $V-K\,\simlt\,1.5$, colors which are 
inconsistent with evolved stars.
Toward the SE, $V-I$ and $V-K$ are redder, which, together with $J-H$,
could suggest the presence of an evolved stellar population.
However, because the $J-H$ structure of the red regions is compact and 
different
from the diffuse structure seen in other colors, it is likely that
dust extinction also contributes to the red colors (see Sect. \ref{gas+dust}).
A reddening of the $J-H$ color by 0.1 mag corresponds to an $A_V$ 
of order unity.

The $H-K$ image (right panel of Fig. \ref{fig:nir}) shows
an even more inhomogeneous distribution.
While the region to the E of the NW cluster has a relatively blue $J-H$,
its $H-K$ color is very red ($\sim$ 0.6, dark grey contours).
Such a combination of colors can only be attributed to hot dust 
(with a temperature $>\,600$\,K) or ionized gas,
since dust extinction or red stars would result in a red $J-H$ as well.
We tend to exclude the hypothesis that the red $H-K$ is due to hot dust
since the UV flux reported by \cite{vidal} is not sufficient to heat
dust to the necessary high temperatures even at 1\,pc radius. 
Indeed, the colors of all regions with red $H-K$ are almost certainly
due to ionized gas, with the exception of the region to the E of
the SE cluster.
There, $J-H$ is also red, which could be due to either red
stars or dust extinction.

The NIR colors of the cut (Fig. \ref{fig:cut}) 
also reflect the information in the color images.
Along the cut, $J-H$ is relatively constant, while
$H-K$ has a local peak ($\sim\,0.5$) in the ``intermediate''
region between the two main star clusters.

\subsection{Hybrid Colors \label{hybridcolors}}

The colors in the optical-NIR hybrid $B-H$ and $V-K$ images (Fig. 
\ref{fig:hybrid}) also show an inhomogeneous spatial distribution, 
with several compact regions being 
significantly redder than their surroundings.
In particular, red $B-H$ ($\ge\,$2.2) is confined to six small clumps
outside the main NW and SE star clusters, and to the SE of the SE cluster.
The same is true for $V-K$ where particularly red colors ($\ge\,1.5$)
are found in small clumps, most notably SE and E of the SE star cluster. 
The NW and SE star-forming complexes are both bluer in $B-H$ and 
$V-K$ than the region between them.

The elliptically-averaged profiles (Fig. \ref{fig:pro}) also show the
effects of small-scale inhomogeneity, as well as large-scale color
changes.
The $V-I$ color of the extended emission is bluer than the main
body, while $B-H$ and $V-K$ are redder; such a trend would be expected for 
ionized gas emission.
The patchy distribution of the hybrid colors can also be seen in
the right panels of Fig. \ref{fig:cut}.
The bluest $B-H$ and $V-K$  colors are observed near the 
NW and SE brightness peaks, with both colors becoming redder farther
away.

\subsection{The C Component \label{ccomp}}

Taken globally,
the NIR colors of the C component  are redder in $J-H$ than those
in the main body, but bluer in $H-K$.
The optical magnitudes of the C component
are $B\,=\,19.19$ and $V\,=\,19.20$ mag and its $V-I$ color is 0.12
\citep{papa02}.
With $H\,=\,17.89$ and $K\,=\,17.97$ for the C component, this gives
$B-H\,=\,1.30$ and $V-K\,=\,1.23$ (1.18 and 1.13 respectively, after the
Galactic extinction correction).
These colors are substantially bluer (in $H-K$, $B-H$, and $V-K$) than those 
in the extended region
around the main body, but similar to those in the SE cluster.

\section{The Age of the Stellar Populations in \izw \label{age}}

A substantial amount of work has been done to reconstruct  
the star-formation history (SFH) in 
\izw\  \citep{dufour,ht95,martin,it98,aloisi,ostlin,legrand00,legrand00p,recchi},
and set limits on the age of its oldest stars. 
As discussed in the Introduction, there is no consensus, and the possibilities
range from a low level of continuous star formation starting some 10 Gyr ago
\citep{legrand00p}, to a first 
star formation episode occurring several Gyr ago \citep{ostlin}, 
to an instantaneous burst of age $\sim$\,15~--~100\,Myr \citep{dufour}.
We examine here whether our deep NIR photometric data can put 
more stringent constraints on these different scenarios and on the age  
of the stars in \izw.

\subsection{Fitting the Broadband Colors: Models and Methods \label{models}}

In accordance with previous work \citep{aloisi,ostlin,recchi},
we consider a star-formation scenario consisting of two distinct episodes:
a recent one which accounts for the young stellar population, and a prior 
one responsible for the older stars.

Both young and old stellar components were modelled with 
synthetic stellar spectral energy distributions (SEDs)
produced using the galactic evolution code PEGASE.2 \citep{fioc} for
an instantaneous burst, with ages ranging between  
0 and 10 Gyr, and a heavy element mass fraction $Z$ = 1/50 $Z_\odot$.
PEGASE.2 uses mainly the ``Padova'' stellar tracks 
as described in the PEGASE documentation; the tracks for $Z\,=\,0.0004$
were published by \citet{fagotto}.
We adopted an initial mass function with a Salpeter 
slope ($\alpha$ = --2.35), and upper and lower mass limits of 120 
$M_\odot$ and 0.1 $M_\odot$.
Synthetic SEDs for continuous star formation were then calculated by
integration of instantaneous burst models over the specific time interval.

The ionized gas emission contribution to the total brightness is important in 
some regions in \izw. 
To calculate the gaseous continuum SED region by region, the observed
H$\beta$ flux and the electron temperature are derived from the optical
spectra \citep{izotov99}. Then the gas continuum is calculated from
the contribution of free-bound, free-free, and two-photon continuum emission 
for the spectral range from 0 to 5 $\mu$m \citep{aller,ferland}. 
Observed emission lines are superposed on the gaseous continuum
SED with intensities derived from spectra in the spectral range 
$\lambda$3700 -- 7500 \AA. Outside this range, the intensities of emission 
lines (mainly hydrogen lines) are calculated from the 
extinction-corrected flux of H$\beta$.

The predicted colors of both stars and gas are obtained by convolving the 
theoretical SEDs with the appropriate filter bandpasses.
The transmission curves for the $B$, $V$, $I$, $J$, $H$ and $K$ bands 
are taken from \citet{bessell90} and \citet{bb88}. The zero points 
are from \citet{bessell98}. 

\begin{figure}
\plottwo{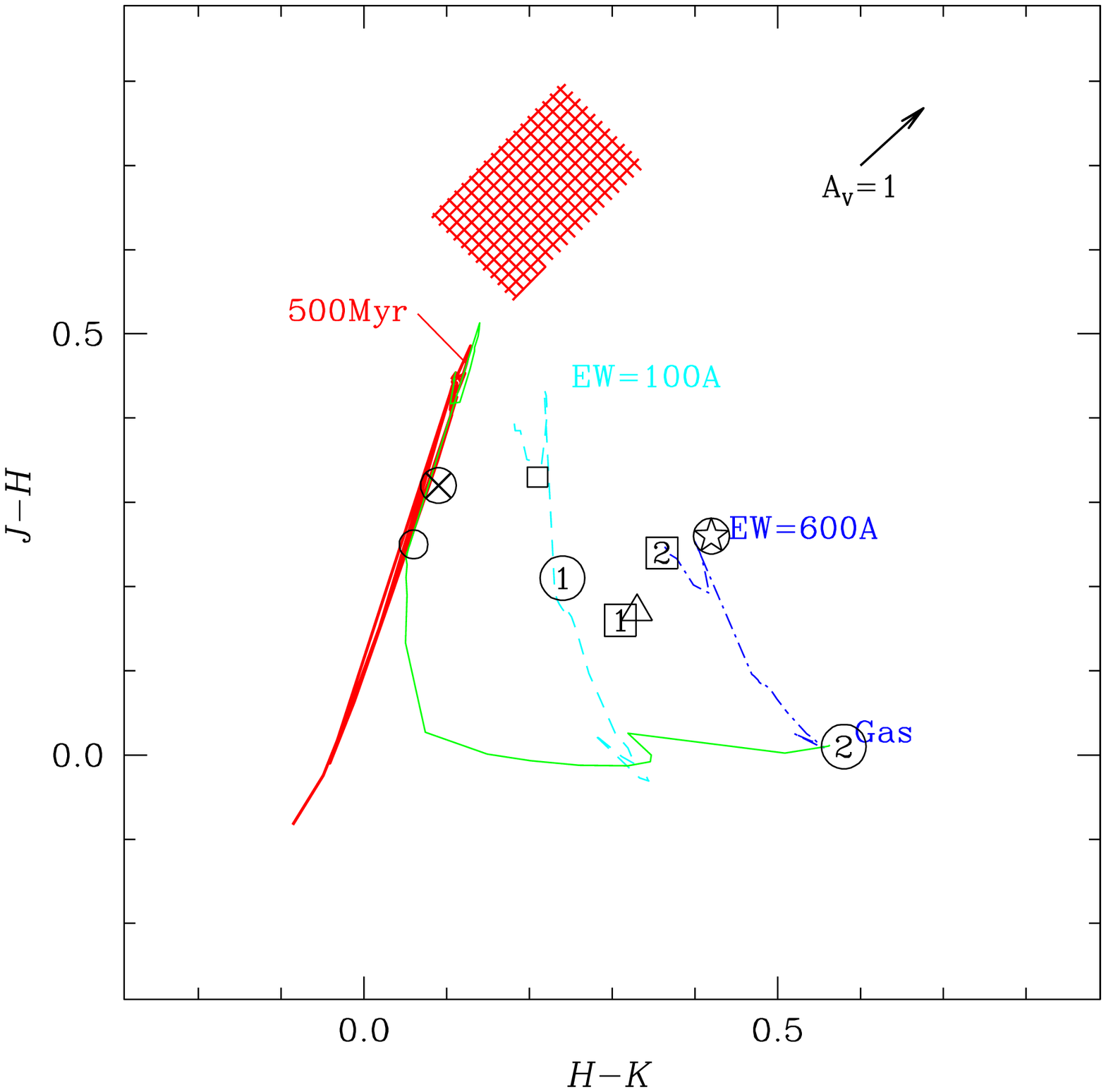}{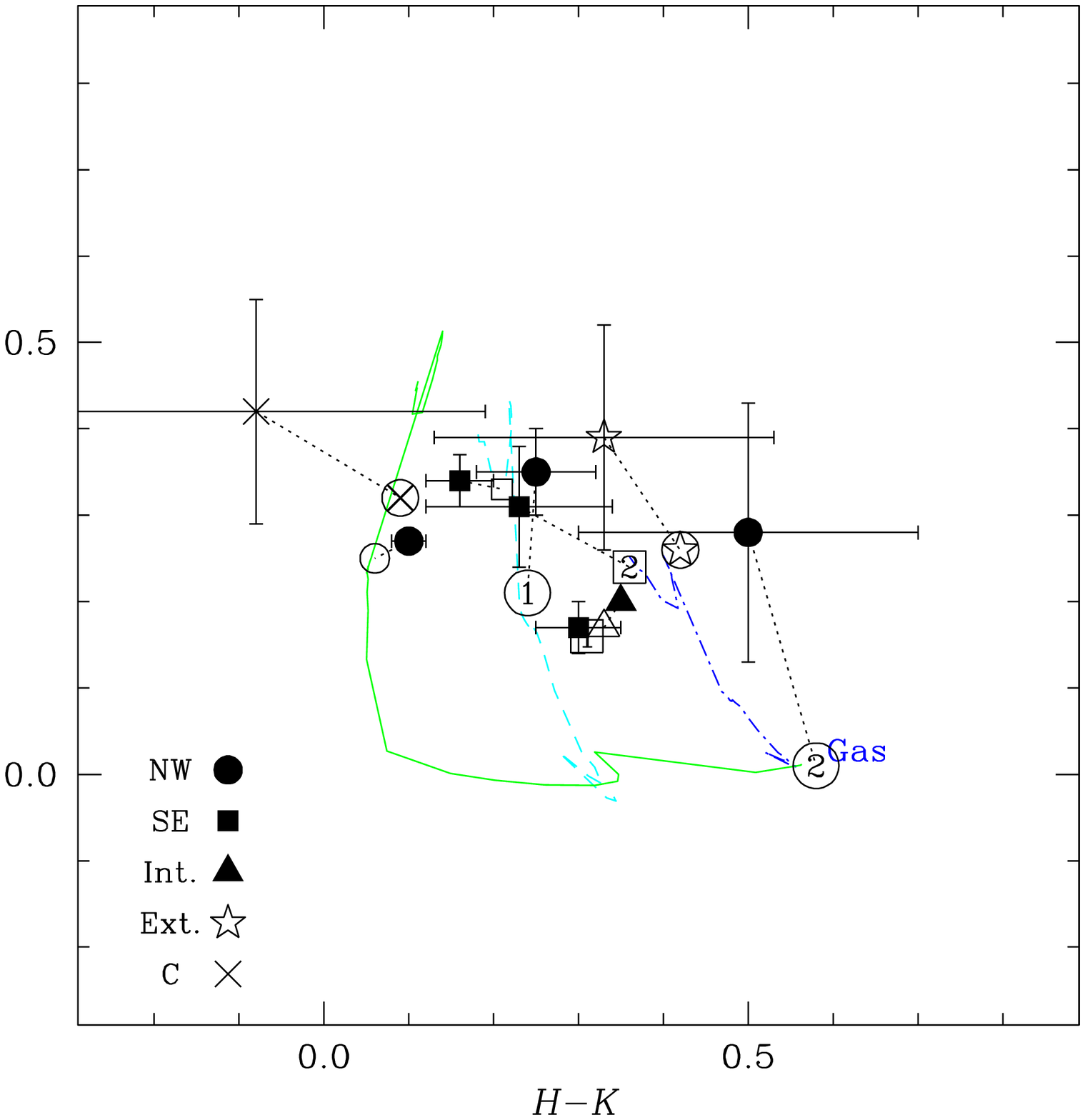}
\caption{$J-H$ vs. $H-K$ plot of \izw\ for colors of the regions
for which we performed aperture photometry, together with
the fitted colors.
Regions are labelled by their designation: NW2, NW1, NW (circles); 
Inter. (filled triangle); SE, SE1, SE2 (squares); 
Exten. (star); and (global) C component ($\times$).
The model colors (shown in both panels)
are denoted with open symbols (and labelled by number
when appropriate), and the observed ones (right panel) with filled symbols;
observed and fitted colors are connected with a dotted line.
Three evolutionary synthesis tracks are shown as described in the text.
Pure gaseous colors are indicated by ``Gas'' (blue).
An arrow indicating an extinction $A_V\,=\,1$ is shown in
the upper-right corner of the left panel.
In the left panel, continuous-burst stars-only colors are shown by a
heavy (red) line; an age of 500\,Myr is marked.
Also shown in the left panel as a grid are the NIR colors of normal spirals,
taken from \citet{roelof}.
\label{fig:m2}}
\end{figure}

\begin{figure}
\plottwo{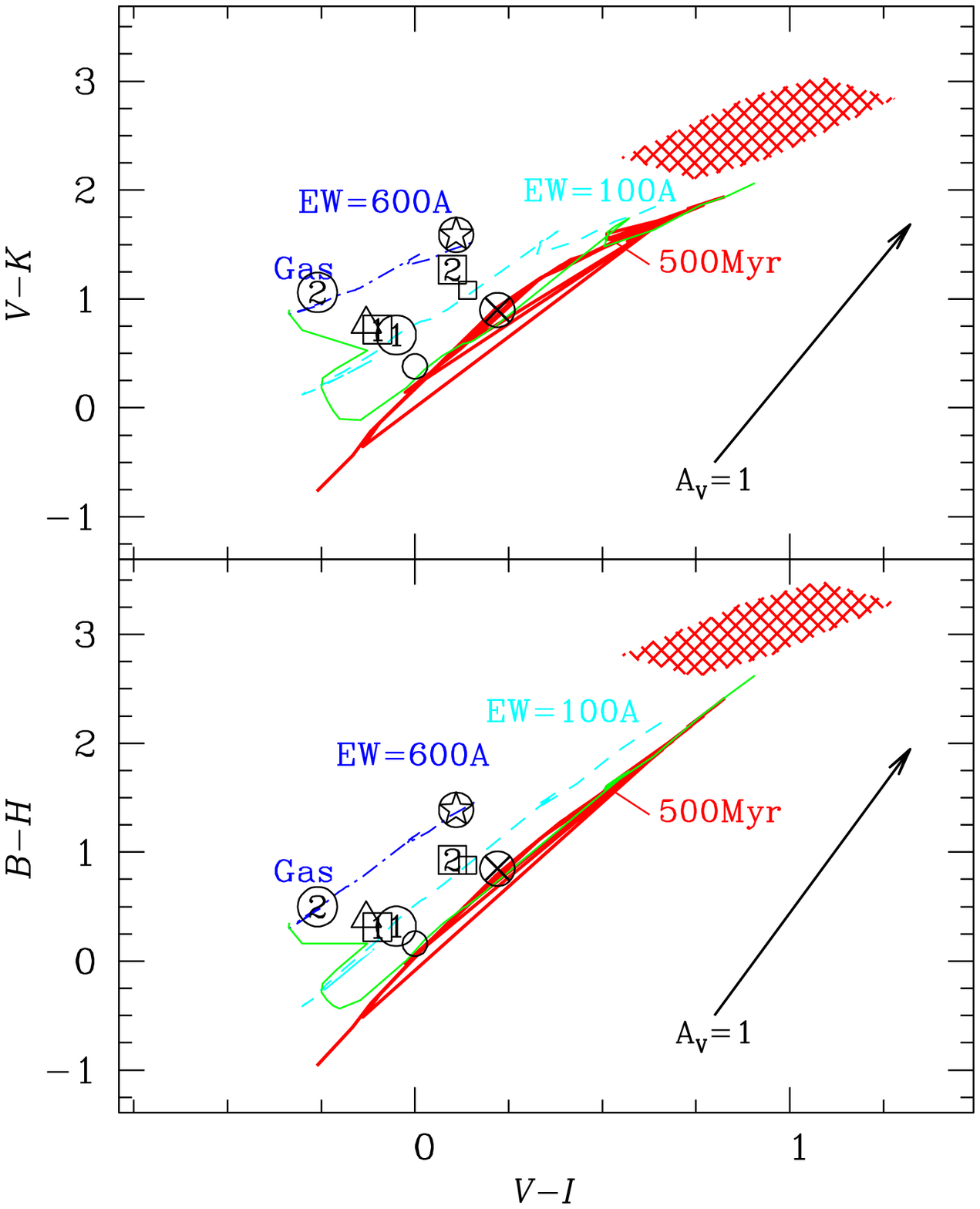}{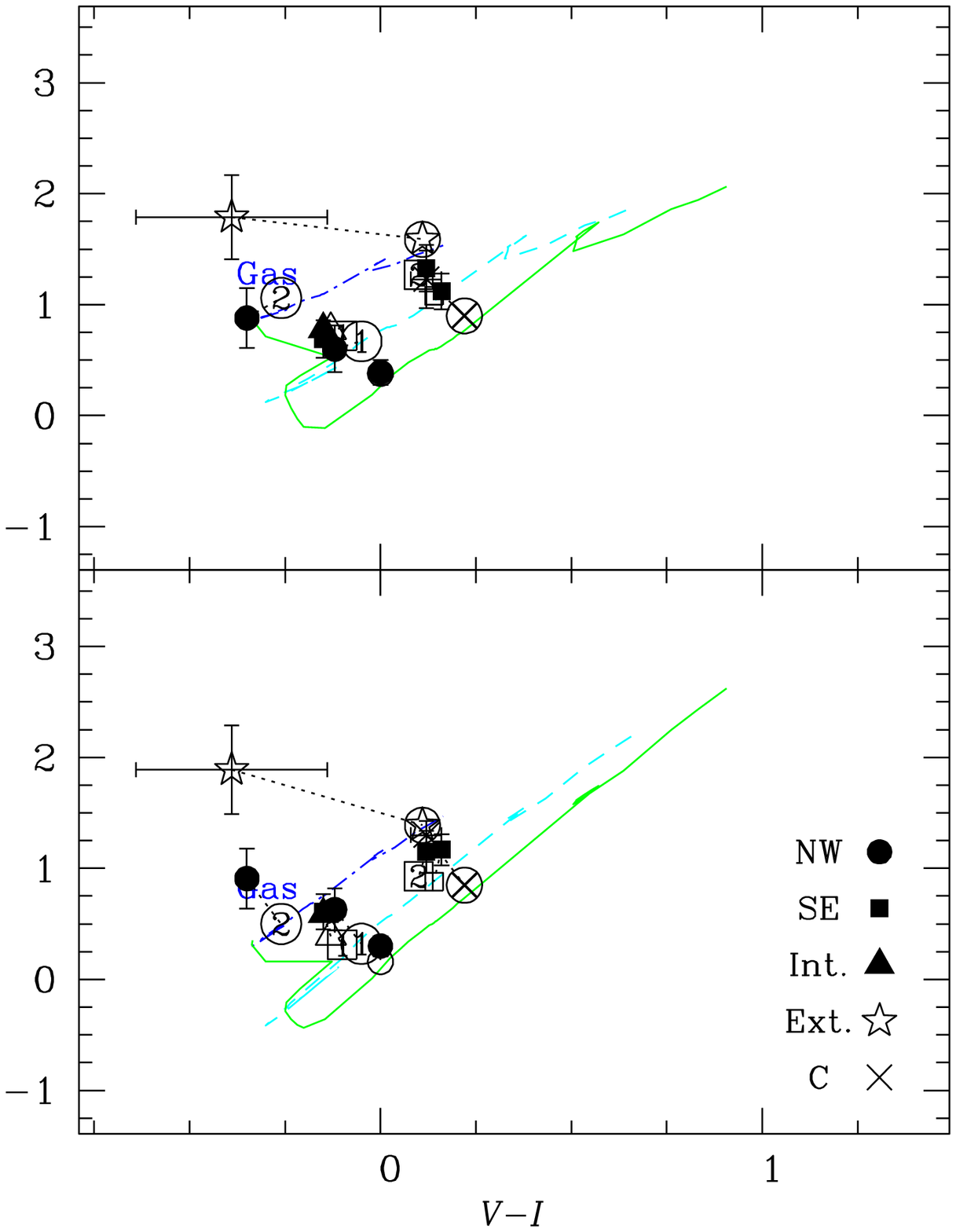}
\caption{Hybrid NIR-optical color-color plots of \izw\ of the regions
for which we obtained aperture photometry, together with the fitted colors.
As in Fig. \ref{fig:m2},
regions are labelled by their designation: NW2, NW1, NW (circles); 
Inter. (filled triangle); SE, SE1, SE2 (squares); 
Exten. (star); and (global) C component ($\times$).
Models are as in Fig. \ref{fig:m2}.
The model colors (shown in both panels)
are denoted with open symbols (and labelled by number
when appropriate), and the observed ones (right panel) with filled symbols;
observed and fitted colors are connected with a dotted line.
Arrow indicating 1.0 visual magnitudes of extinction are shown in
the left panels.
\label{fig:m4}}
\end{figure}

Generic colors of the young population only (instantaneous burst)
are shown in the left panels of Figs. \ref{fig:m2} (NIR colors) and \ref{fig:m4}
(optical and hybrid).
Gas emission is taken into account in these curves by adding  
the {\it observed} gaseous SED 
from optical spectrophotometric observations of \citet{izotov99} to the calculated 
stellar SED of the younger population; its 
contribution is determined by the ratio of the observed equivalent width (\ew)
of the H$\beta$ emission line to the one expected for pure gaseous emission.
The color-color tracks shown in Figs. \ref{fig:m2} and \ref{fig:m4}
are those predicted by our instantaneous burst models for stellar
populations with ages from 0 to 10 Gyr and for 1/50\,$Z_\odot$ metallicity,
with line and continuum gaseous emission added. 
The relative intensities
of the emission lines are taken to be the same as in the NW component.
The short-dashed (cyan) line shows populations with fixed \ew(H$\beta$)\,=\,100\AA, and 
the dot-dashed line (blue) populations with \ew(H$\beta$)\,=\,600\AA.
The light solid (green) line shows a scenario in which \ew(H$\beta$)
changes self-consistently over time, namely the gaseous emission is
defined by the ionizing flux of the stellar population;
an age of 500\,Myr is indicated with an arrow, with older ages being
toward the red in both colors.
These tracks differ from those calculated by \citet{sb99} 
mainly because they are characterized by a lower metallicity (1/50 $Z_\odot$), 
and take line emission into account.
Pure gaseous emission is denoted by ``Gas''. 
Colors of 
continuous-burst stellar populations alone (without gas) are shown by heavy solid (red) lines.
There are more than one of these (evident in Fig. \ref{fig:m4}) because
we considered several initial and final burst ages; in the NIR these differences
are insignificant, while in the optical$+$hybrid they are more pronounced.
Also shown in the left panels as a grid are the NIR colors of normal spirals,
taken from \citet{roelof}.

\placefigure{fig:m2}

\placefigure{fig:m4}

The temporal evolution of the self-consistent model shows clearly the
epoch where gas ceases to dominate the SED, since there is a sharp shoulder
in the track, and the continuous-burst stars-only models converge with it. 
Red $H-K$ and blue $J-H$ and $V-I$ correspond to the youngest 
ages. 
There is an ambiguity in the hybrid colors since red $V-K$ (and to a lesser
extent $B-H$) can be due to either gas or evolved stars.
We will exploit the five independent colors at our disposal to resolve
this ambiguity.

The observed color-color diagrams for the seven regions in the main body, the extended
region around the main body, and the C component are shown in the
right panels of Figs. \ref{fig:m2} (NIR) and \ref{fig:m4} (optical$+$hybrid). 
It is evident that the observed NIR and hybrid optical-NIR 
colors can be well reproduced by the models. 

\subsubsection{Constraints on the Old Stellar Population}

To reconstruct the star formation history in I Zw 18 from its observed
broadband colors, we proceed in the following manner. We first
search the grid of synthetic models for the one 
which fits best the five broadband colors 
($V-I$, $B-H$, $V-K$, $J-H$, and $H-K$) observed for the seven positions of
aperture photometry in the main body of I Zw 18, the extended region around
it, and the C component.
In all cases, we modelled the old stellar population with a continuous
episode of star formation, but
considered both instantaneous and continuous scenarios for the young stellar
population. 
The old stellar population, by definition, is composed of 
stars formed prior to the stars composing the young stellar population.

We derived the colors expected in each observing aperture by
``mixing'' curve calculations, usually used in connection with color-color diagrams.
In such calculations, one starts with a ``main'' color associated with
a physical component, in our case young instantaneous- or continuous-burst 
stellar populations. 
That color is then ``mixed'' with colors of other physical components,
in our case continuous-burst older stars
and gas emission (including both continuum and line emission as discussed above). 
Finally, foreground extinction is
taken into account by reddening the ``mixed'' colors.
Mathematically,
the resulting colors do not depend on which component defines the ``main'' color.

For each region, we derived the model colors
($V-I$, $B-H$, $V-K$, $J-H$, and $H-K$) that best fit the observed
ones by minimizing the $\chi^2$.
The uncertainties in the $\chi^2$ are the photometric uncertainties in the colors,
obtained by adding in quadrature the photometric uncertainty of each filter.
Color errors were constrained to be at least 0.05\,mag 
because of possible systematic problems with \hst/ground-based transformations. 
Free parameters in the model fitting
include the age of the young stellar population and its burst mode of
star formation (instantaneous or continuous), 
the age of the old stellar population (constrained to be a continuous
burst),  the $J$ luminosity fraction of old stars $r_*(J)$, 
the $J$ luminosity fraction of ionized gas $r_{\rm gas}(J)$,
and the foreground visual extinction $A_V$.
Rather than fixing $A_V$ to the values derived from the Balmer decrement,
we let $A_V$ vary in the fits, since ionized gas may be subject to a different
extinction than the stellar emission \citep{calzetti}.

However, fits performed in this way
are unconstrained since we have five colors and five
free parameters (six if the young-population burst scenario is included). 
Thus, to limit the parameter space to be explored,
we had to impose additional observational constraints. 
One such constraint is the \ew s of the H$\alpha$ line
at different locations in \izw\  \citep{izotov01} 
which limits the range of ages the young stars can have. 
Furthermore, photometric studies of BCDs  
[e.g., \citet{loose}; \citet{papa96}], have shown that  
the same component of evolved stars is present in all regions within a 
galaxy. 
Therefore, we constrain the evolved stellar population to be the
same everywhere in \izw.
To determine which ages are possible,
we need to examine the colors of a region relatively free 
of the ionized gas emission known to be widespread in some parts of 
\izw\  \citep{izotov01} and which contaminates the  observed 
broadband colors. 
The C component is such a region;
it has the lowest \ew(H$\beta$) of the \izw\ complex. 
Therefore, in principle, we can first fit the C component colors,
without having to worry about gas contamination;
then, the old stars in the \izw\ main body
can be represented by the best-fit older stellar population in 
the C component. 
Such a procedure is supported by 
optical spectroscopy and imaging which suggest that star formation 
is oldest in the C component  (age $\sim\,$100~--~200\,Myr) 
and has propagated into the main body \citep{ht95,dufour,it98}.

We therefore fit the C component by setting the gas fraction $r_{\rm gas}(J)$
and extinction $A_V$ to zero, and letting the age of the young and old populations
vary, together with the young-star burst scenario and the old-star fraction $r_*(J)$.
By constraining the model prediction of the  H$\alpha$\,\ew\ in the burst scenario
of the young population to be consistent with its observed value 
(20~--~60\AA\ \citep{izotov01}),
we find the colors of the C component to be best fit
by a young instantaneous-burst (IB) stellar population 15\,Myr of age, 
with a dominant $r_*(J)\,=\,0.6$ older stellar population of age
100--200\,Myr. 
The second-best fit was obtained with a young continuous-burst (CB) 10--20\,Myr
of age, combined with $r_*(J)\,=\,0.6$ of 100--200\,Myr older stars.
The third-best fit was similar to the best one, but with $r_*(J)\,=\,0.5$
of a 20\,Myr young population.
These results are reported in Table \ref{table:fitc}. 

The minimum $\chi^2$ value \chimin\,=\,16.7 is obtained in the IB scenario,
and gives a best fit with
root-mean-square (RMS) residuals (over the five colors) of 0.21 mag, 
while the CB \chimin\,=\,17.0 corresponds to an RMS of 0.23 mag.
If we consider all the old stellar populations found in solutions with
a $\chi^2$ value between \chimin\ and
\chimin\,$+$\,1, we find that, in addition to the best-fit range of 100--200\,Myr
age, plausible other age ranges for the old stellar populations for \izw\
include
100--500\,Myr, 10--200\,Myr, 10--500\,Myr, and 3--500\,Myr.
Although stars older than this gave substantially worse fits,
to be conservative, and because the age of \izw\ is so controversial,
we have included in the subsequent analysis the following
additional older stellar populations:
3\,Myr--1\,Gyr, 100\,Myr--1\,Gyr, 0--1\,Gyr, 100\,Myr--2\,Gyr.
As a result, for the remaining main-body regions, we have considered
a total of nine possible old stellar populations.
Populations older than 2\,Gyr are ruled out, since
the \chimin\ values associated with them (together with the \ew\ constraint on
the young stars) were ranked at $\sim$ 600$^{\rm th}$ in the best
(IB) case. 
The worst of the older populations retained for further
consideration (e.g., 2\,Gyr) ranked 263$^{\rm rd}$.

\subsubsection{The Young Stellar Populations, Gaseous Emission,
and Extinction}

We then fit the colors of each region in the main body of \izw\
by letting the age of the young stellar component vary, 
subject to \ew\ constraints, 
as well as the parameters $r_*(J)$, $r_{\it gas}(J)$, and $A_V$.
The old age and the burst scenario of the young population
(IB or CB) was constrained to be the same for all regions.
This was done for
each of the ten different possible old stellar populations as defined above.
We let each fraction in the mixing calculation vary from 0 to 1 in 10\% increments,
and let $A_V$ vary from 0 to 1, in increments of 0.1\,mag.
Therefore, considering all the possible young stellar populations
(63), together with the nine candidate older populations, we calculated for each
region 567,000 ($63\times 9\times10^3$) sets of colors;
with the two burst scenarios, we have considered more than a million color sets. 
\chimin\ summed over all five colors for each region was chosen from this ensemble. 
The global \chimin, \gchimin, for a specific older stellar population and younger
burst type was defined by summing over all regions (including the C
component) the \chimin\ for each region at the particular older age (and 
younger burst type).
The results are shown in Table \ref{table:fit}, in order of increasing \gchimin.

The optical and hybrid observed within a 2\arcsec\ 
aperture are plotted together 
with the fitted colors in Figs. \ref{fig:m2} and \ref{fig:m4}.
In both figures, the best-fit colors are shown in both panels, and
connected to the observed ones in the right panel with a dotted line. 
The fitted values for the five colors for each region can be seen to be
consistent with those observed, within the observational errors. 

\subsection{Results}

\begin{figure}
\plotone{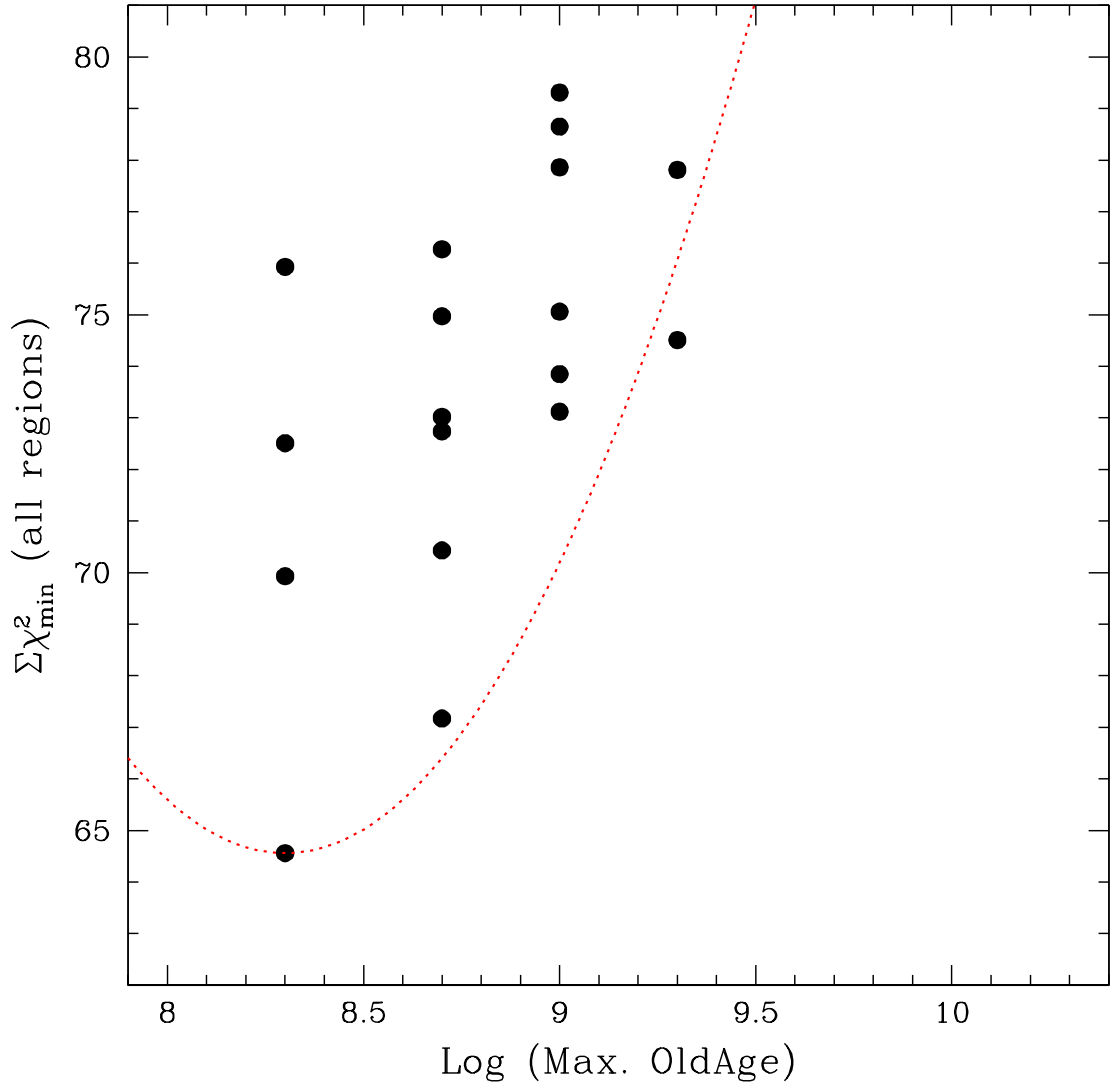}
\caption{
The sum $\Sigma\chi^2_{min}$ over all regions plotted as a function
of the logarithm of the maximum age in the older stellar population; only the 18
lowest values are shown.
The two star-formation-history scenarios are not distinguished in this figure, nor are
the younger ages (either for the younger population, or for the lower
limit in the oldest). 
The dotted curve shows the parabola which best approximates the envelope
of lowest \gchimin\ values.
\label{fig:chisum} }
\end{figure}

The best fit as given by the minimum \gchimin\ in
Table \ref{table:fit} is obtained for an underlying 
old stellar population of age 100--200\,Myr. 
The \gchimin\ values are plotted as a function of old population age in Fig. \ref{fig:chisum},
where only the 18 lowest values are shown;
these values are also reported in Table \ref{table:fit}. 
The age in the plot is in fact the oldest age in the older stars, since there is 
ambiguity because of the numerous starting and ending CB times for the
old burst in our model grids.
Inspection of Fig. \ref{fig:chisum} shows the clear minimum for an older age of
200\,Myr.
Although the difference in $\chi^2$ between an oldest age of 200\,Myr and
500\,Myr is not significant enough for us to definitely exclude 500\,Myr
as the oldest age, we can definitely exclude an upper limit of 1\,Gyr.
In fact, the best fit with an oldest age of 1\,Gyr is ranked 8$^{\rm th}$,
and has $\Delta\chi^2\,=\,8.5$ with respect to \gchimin.
As for the burst scenario, our fits 
tend to suggest that an instantaneous burst rather than a continuous one
is more likely for the younger population,
since seven of the nine lowest \gchimin\ values are for IBs.

Table \ref{table:fit} shows that, with two exceptions, the lowest \gchimin\
also corresponds to the lowest regional \chimin.
Hence, our requirement that the same 
evolved stellar population be present in all regions is not unduly
constraining our results, since the condition is met naturally by our fits. 
One exception, NW2, is insignificant, since according to our fits
it contains no evolved stars; consequently the resultant \chimin\ is insensitive to older age.
The other exception (SE) reflects the difficulty of fitting
a complex region with a simple model.
The SE component shows the most spatially variable colors in the entire main body,
and undoubtedly suffers from the effects of dust extinction.

The parameters of the best-fit model with an older population of age
100-200\,Myr are given in Table \ref{table:par}.
In all cases, the RMS residuals of the fits (Col. 6, 1st line) are smaller than
or comparable to the RMS photometric uncertainties (Col. 6, 2nd line),
suggesting that our fits are rather good approximations to the colors
(see also Figs. \ref{fig:m2} and \ref{fig:m4}).
Inspection of Table \ref{table:par} shows that 
the C component and the main SE cluster apparently host the highest fraction 
of evolved stars with $r_*(J)$\,=\,0.6.
The other regions have lower $r_*(J)$, with zero old stars in the farthest
region to the NW (NW2); indeed 80\% of the emission in NW2 is ionized gas.
In the main body,
the old-star fractions suggest that star formation has been proceeding the longest
in the main clusters, SE and NW, since half or more of the light there can be
attributed to a 100-200\,Myr continuous burst.

The ages of the younger burst range from 3\,Myr in the NW part of the main body
to 10\,Myr in most of the SE (with the exception of SE1 which has an age of 3\,Myr).
In the C component, the younger burst is older, with an age of 15\,Myr or more
(see also Table \ref{table:fit}).
Because of the gas emission,
in most of the regions in the main body (with the exception of the NW cluster), 
young stars comprise less than half of the observed flux.
The highest fraction of young stars is present in the NW star cluster (50\%),
while toward the SE, the fractions range from 10\% to 30\%.
These results will be compared with previous work in $\S$\ref{comparison}.

\subsection{Gas Contamination and Dust Reddening\label{gas+dust}}

We discuss here the evidence that the emission in many 
regions in \izw\ is dominated by gas and dust, rather than by old stars, 
and how neglecting this can affect the derived age.

\subsubsection{Spatial Inhomogeneity of the Interstellar Medium}

Table \ref{table:par} shows that in the main body only the bright NW cluster
is free of ionized gas.
With the exception of the main NW cluster, {\it all} of the remaining regions 
contain a significant amount of ionized gas.
As mentioned above,
the gas fraction $r_{\rm gas}(J)$ is highest to the NW (NW2), where 80\% of
its light is due to gas emission.
This gas contribution is lowest in the SE, consistent with the relatively
high fraction of evolved stars derived.

These results are compatible with the qualitative analysis in $\S$\ref{optcolors}
and with other data. 
Deep Keck spectra of \izw\  obtained  
by \citet{izotov01} reveal extended gas emission around the NW 
component, with H$\alpha$ emission detected as far as 30\arcsec\ from it.
The situation in the NW star cluster is contrary to this, since
that region is only well fit by pure stellar emission, 
without any gas contribution.
Our finding is also consistent with the H$\alpha$ map of \izw\ 
which shows an apparent ``hole'' in the gas distribution at the 
location of the NW component \citep{cannon}. 

Extinction is also measurable in the main NW and SE clusters ($A_V\,=\,0.1$\,mag),
and in the extended region.
According to our fits, the light in the extended region surrounding the main
body of \izw\ contains 40\% evolved stars of age 100-200\,Myr, and 60\% gas,
together with an extinction $A_V\,=\,0.2$ mag. 
The high $\chi^2$ and the large uncertainties\footnote{As mentioned in $\S$\ref{cuts},
these are the standard deviation over the ellipse.} here reflect the difficulty of 
using an elliptical average, especially because  in the NW2 region, the
emission is 80\% gas without a measurable contribution from an evolved
stellar population, while in the SE, there appear to more old stars
(40\% of the light) and less gas.

In the SE region, $V-I$ is redder ($>$\,0), as are the other colors,
and, from Table \ref{table:fit}, it appears that the 
color gradient evident in Fig. \ref{fig:cut}
is due to a population gradient, coupled with some extinction and
a spatially-varying contribution from ionized gas.
A greater extinction in the SE cluster is supported also by other data
\citep{it98, cannon}:
small-scale $A_V$ from the Balmer decrement in the SE region is 0.24 mag as compared to
$A_V$ = 0.04 mag for the NW region.
The slight difference between these and the results of our fits is almost certainly due
to different aperture sizes and the spatially-varying interstellar medium (ISM).

Indeed, using the \hst/WFPC2 extinction maps, \citet{cannon} show
that the H$\alpha$/H$\beta$
ratio varies considerably spatially, and show that such extinction can
be produced by $2-5\times10^3$\,$M_\odot$ of dust.
The structure in the extinction inferred from the recombination line ratios is
very similar to the one seen
in our optical/hybrid color images: the SE and the E
side of the intermediate zone which are the most affected by extinction
(see Figs.\ref{fig:opt}, \ref{fig:hybrid}) are also the sites of the reddest colors.
Cannon et al. speculate that there is a band of extinction between the NW
and SE components, but are unable to measure it because of the absence of
optical line emission.
Our color images, in particular $B-H$, reveal such a band, suggesting
that much, if not all, of the structure seen in the colors is due to
variations in the ISM, rather than to gradients in the stellar populations. 

\subsubsection{Color structure\label{trans}}

The {\it structure} of the extended emission and the color gradients can
also help understand its nature.
As discussed in $\S$\ref{broadband},
the colors in \izw\ vary on small spatial scales.
Such variations are not the smooth ones that would be expected for stellar 
population age gradients, but rather are patchy and inhomogeneous, 
common signatures of variations in dust extinction and gas emission. 
Indeed, the reddest structures in the colors are associated with  
compact sources rather than with smooth sheets of underlying red stars,
which are conspicuously absent in our images.

\begin{figure}
\plottwo{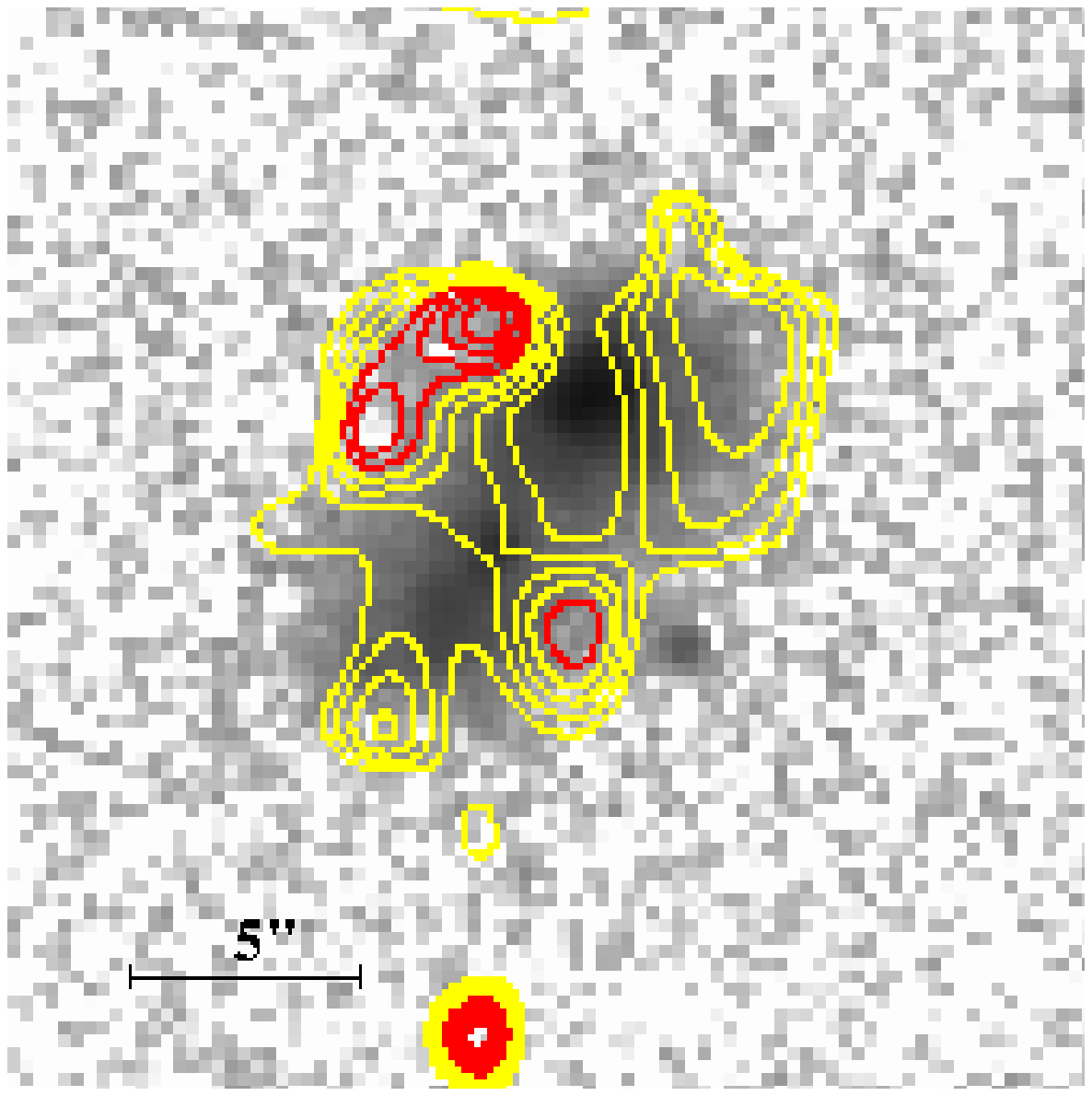}{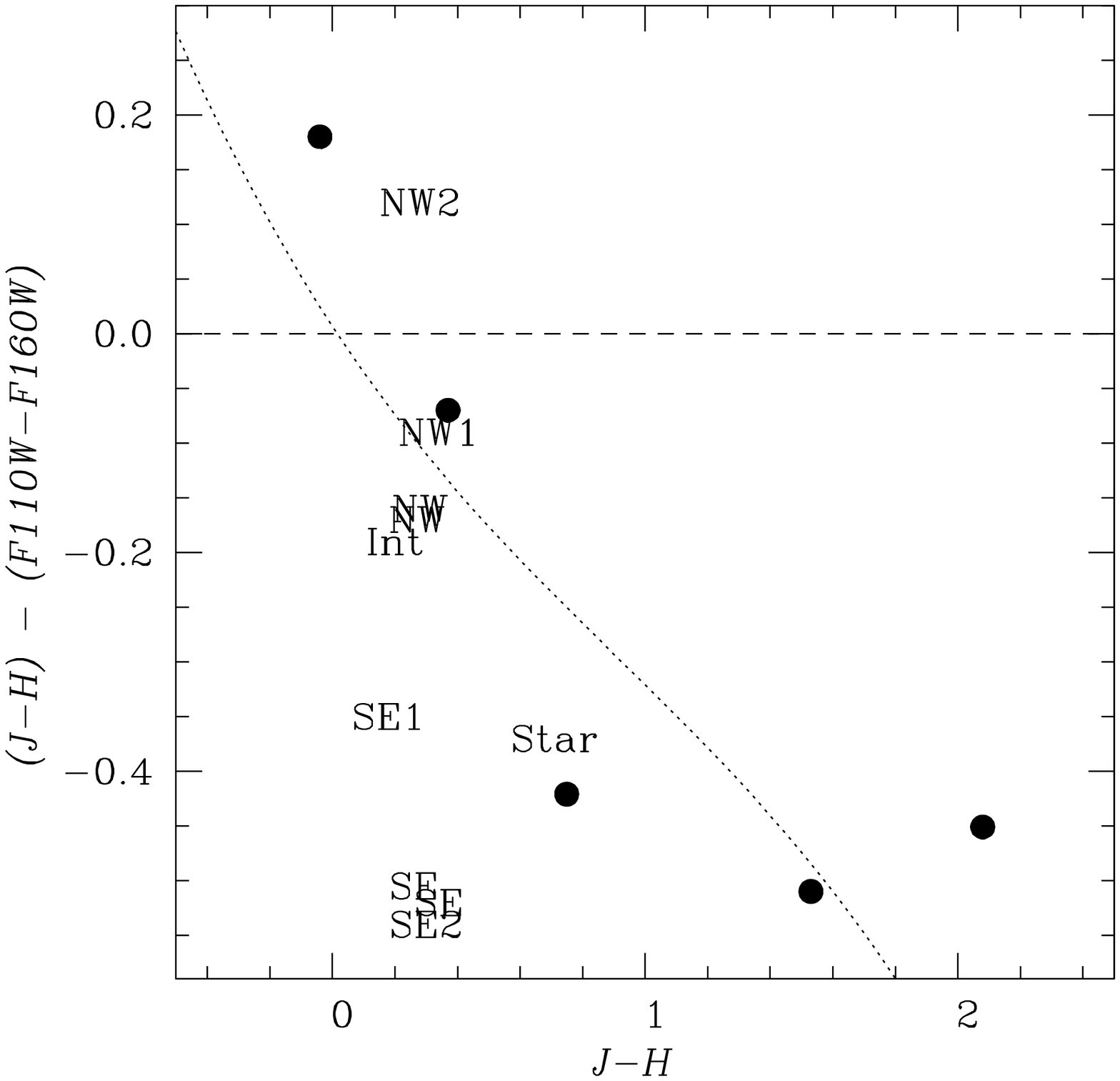}
\caption{The difference in the color transformation 
for \hst\ and ground-based photometry. Left panel:
difference $(J-H)-$trans({\sl F110W}--{\sl F160W})
of the ground-based $J-H$ images of \izw\
and the transformed NICMOS {\sl F110W}--{\sl F160W};
right panel: the same color transformation for the individual
regions in \izw, individual stars 
taken from {\sl HST}/NICMOS web page (filled circles), and for 
a field star (``Star'') in the \izw\ image.
The transformation of the {\sl F110W}--{\sl F160W}\  color was performed
according to \citet{origlia}, and the (negative) difference between it
and ground-based $J-H$ is shown in the right panel as a dotted line.
The horizontal line shows zero difference in the \hst\ and ground-based colors.
The contours (left panel) and the negative magnitudes (right)
correspond to a negative difference, 
in the sense that ground-based $J-H$ is {\it bluer} than
transformed {\sl F110W}--{\sl F160W}. 
In the left panel, {\it dark grey [red]} contours correspond to
$(J-H)-$trans({\sl F110W}--{\sl F160W})\,$<\,-0.5$, and
{\it white [yellow]} to
$-0.5\,<\,(J-H)-$trans({\sl F110W}--{\sl F160W})\,$<\,-$0.1.
\label{fig:trans}}
\end{figure}

Another signature of the ISM in \izw\ can be found from the comparison of the
ground-based $J$ image with the \hst/NICMOS {\sl F110W}\ image. 
The difference $(J-H)$ -- transformed  ({\sl F110W}--{\sl F160W}) 
(see Sect. \ref{hst})
is shown in the left panel of Fig. \ref{fig:trans},
with contours overlaid on the ground-based $J$ image.
If the analytical transformation of \citet{origlia}
holds, we would expect a difference
between transformed {\sl F110W}--{\sl F160W}\ and $J-H$ to be close to zero;
that is the case for the NW and ``Inter.'' regions of the main body.
However, to the E of the intermediate region, to the SE and to the W of 
the NW star cluster, and to the S, the difference can be very negative.
The highest (dark grey) contours correspond to a difference of $\simlt\,-0.5$\,mag, in 
the sense that the transformed {\sl F110W}--{\sl F160W}\ is redder than the ground-based $J-H$.

\placefigure{fig:trans}

The failure of the analytical color transformation 
is seen perhaps more clearly in the right panel of Fig. \ref{fig:trans},
where we have applied the color transformation to individual calibration stars, 
the seven discrete regions in the main body, and a field star in the image field-of-view.
The field star is labelled ``Star'', but is probably a galaxy \citep{ostlin}.
The calibration stars and their photometry (shown by filled circles)
were taken from the \hst/NICMOS web page.
It can be seen from the figure that the NW cluster and the ``Star'' obey the 
nominal transformation rather well, 
but the SE cluster and surroundings are highly discrepant:
the {\it transformed} {\sl F110W}--{\sl F160W}\ is 0.5\,mag redder 
than the true ground-based $J-H$ color.
In NW2, the $J-H$ color is {\it bluer} than the transformed {\sl F110W}--{\sl F160W}\
consistent with slightly different gas fractions in the two sets of colors.

In general, the largest differences are located in the regions with
the reddest colors. 
The spatial structure of these regions in the NIR images is
very similar to that of the hybrid color images. 
These differences 
are good tracers of gas emission and dust extinction because, while 
the color transformation from {\sl F110W}--{\sl F160W}\ to $J-H$ is valid 
for normal stars, it fails where the colors are contaminated 
by other processes.
That the ISM is at least partly responsible for the large 
differences is particularly evident 
in the negative region to the W of the NW brightness peak: this site 
has one of the largest H$\alpha$ equivalent widths \citep{izotov01} 
in \izw.

Gaseous emission (including lines) may not affect ground-based $J-H$ much,
since its $J-H$ $\sim\,0$, although it may make the color slightly bluer. 
However, the sensitivity to extinction of {\sl F110W}--{\sl F160W}\
is 0.3\,$A_V$ \citep{holtzmann}, more than 3 times that of ground-based $J-H$. 
The opposing effects of gas (which makes $J-H$ bluer)
and dust extinction (which reddens the colors) make it difficult to quantify
their combined effects. 
We can only emphasize that if these effects are not taken into 
account in the interpretation of the observed colors, the inferred age for 
the stellar populations may be incorrect. 

\subsubsection{Gas Emission and its Effect on Color Fitting\label{gas}}

To better assess gas contamination and its effect on the fit results,
we have estimated the gas fractions colors in the NW and SE clusters using 
two different techniques:
1)~from the optical recombination line ratios [e.g., \citet{it98}]
and assuming zero extinction, infer the NIR line fluxes and gas continuum;
2)~from our new NIR spectra obtained at the Keck and UKIRT telescopes 
(Hunt et al., in preparation), measure
the NIR recombination lines and infer the gas emission in the photometric filters.
For (1), we have used the recipes of \citet{joy} to derive the
gas continuum emission in the spectroscopic apertures (NW 4\farcs2$\times$1\farcs5,
SE 3\farcs6$\times$1\farcs5).
The extrapolation from the optical gives slightly lower gas fractions 
(15\% in the SE and 5\% in the NW) 
than our NIR spectra (SE: 32\%, NW: 22\%), however the former do not take
line emission into account.
Both strategies give results which are roughly consistent with each 
other, and broadly consistent with the results from the color fits,
especially considering the different aperture sizes and orientations.
The spectroscopic apertures are larger than the photometric ones and unequal (NW is larger than SE),
and the spectroscopic slit was oriented at a slightly different angle (PA\,=\,139$^\circ$)
than our photometric cut (PA\,=\,149$^\circ$).
The larger gas fractions measured by our new NIR spectra relative to the
optical extrapolations may also suggest a larger extinction.
In spite of these caveats,
the important result is that both techniques confirm spectroscopically
that the SE cluster is more contaminated by gas than the NW, corroborating
the results of our fits.

To determine if the age of the evolved stars inferred from our fits 
depends sensitively on our treatment of gas emission, we also ran a series of stars-only fits,
that is to say {\it without including the contribution of gas emission to the model colors}.
Although the resulting region \chimin\ values were much larger (\chimin\ $\times\,3$),
the maximum ages for the older stellar population remain 200-500\,Myr.
Moreover,
these stars-only fits give the same ranking for the older stellar populations:
namely the \gchimin\ was found for an IB of age 100-200\,Myr, with the
next smallest value given by one of age 100-500\,Myr.

To ascertain if the age of the evolved stars inferred from our fits 
depends on {\it which} colors are fit, we repeated the stars-only fits on
different subsets of colors. 
We first eliminated one or both of the hybrid colors $V-K$ and $B-H$
from the fits, since these colors are presumably the most sensitive to red
old populations.
However, the results do not change:
we obtain the same upper limit to the older age of 200-500\,Myr, and
the ranking of the fits is the same as with all five colors

We then eliminated $V-I$ from the fits, and again modelled the four remaining colors 
(now two hybrid ones and two NIR) without including gas emission. 
This was done for two external regions (Exten., SE2) with {\it red} colors ($V-I$, 
$V-K$, and $B-H$) so that 
we could effectively check the case most favorable for evolved stars.
It turns out that
{\it eliminating this purely optical color from the fits significantly changes
the estimate for the old age in these two regions}.
Without $V-I$, the age of the evolved stars becomes 100Myr-2Gyr;
$r_*(J)$ becomes unity for the extended region and 0.7 for SE2
(instead of 0.4). 
The reason for this is that the $V-I$ of the gas is extremely blue ($\sim\,-0.4$) relative
to old stars ($V-I\sim\,+0.4$), while the 
$B-H$\footnote{These colors all refer to $Z_\odot/50$ stellar populations and ionized gas.}
($\sim\,0.8-1.0$) and $V-K$ ($\sim\,1.2-1.3$) of the gas are similar to those
of evolved stars ($B-H\,\sim\,1.3-1.8$, $V-K\,\sim\,1.6-1.8$).
Therefore, hybrid colors can be equally well fit by gas and evolved stars;
red $B-H$ and $V-K$ are {\it ambiguous} since ionized gas and old stars
are similarly red.
Thus they cannot be used 
to discriminate gas from evolved stars in galaxies such as \izw.
$V-I$ on the other hand is a good discriminator since gas and evolved stellar colors
differ by $\simgt\,0.8$\,mag.
It is clear that had we not included gas emission in our fits,
and avoided incorporating $V-I$, we would
have concluded that the stellar populations in \izw\ are 2\,Gyr old.
This age is a factor of four greater than our most stringent upper limit.

\subsection{Comparison with Previous Work on \izw\ \label{comparison}}

We have shown in the preceding sections that the optical, NIR and hybrid 
colors of all regions in \izw\  can be explained by an underlying 
stellar population not older than $\sim$ 500 Myr, when the effects of 
gaseous emission and dust on the colors are taken into account, and when
a full set of colors, including $V-I$, is incorporated in the modelling.
This 
relatively young age is in disagreement with  the findings of \citet{ostlin}
who found a number of red stars in the CMD obtained from \hst/NICMOS images
which he identified as AGB stars. 
Converting the \hst\ {\sl F110W}--{\sl F160W}\ to $J-H$ using the analytical transformation
described in \S\ref{trans}, 
comparing with isochrones from \citet{bertelli}, and taking into account 
photometric errors, \"{O}stlin argued that an age of at least 1~Gyr for 
\izw\  is necessary to explain the NICMOS CMD.
Moreover, an age of $>$ 125~Myr was attributed to the unresolved
stars in the SE complex, with a younger age of 20\,$-$\,100\,Myr in 
the NW region. 

The old age of \izw\ inferred by \citet{ostlin} is based on 23 stars in the NICMOS
CMD with {\sl F110W}--{\sl F160W}$\,\simgt\,0.8$ (see his Fig. 3). 
While our model fits also suggest that the SE component is older than 
the NW one, in agreement with the  findings of \citet{aloisi} (see below),
our data would not reveal such a small fraction of red stars.
In the mean, each of these stars has a magnitude of $\sim$\,23.5\,mag
in the {\sl F110W}\ filter, with the reddest stars being the
faintest.
Summing the luminosity of these red stars, we obtain a total {\sl F110W}\ mag of
20.1 which, assuming $J\,\approx\,${\sl F110W}, is only $\sim$ 2\% of
the $J$ total luminosity in \izw.
Such a small fraction is below the detectability limits of our observations,
because of the worse 
ground-based resolution as compared to that of HST , and because of the low impact on the
colors over scales of $\leq$\,0\farcs5.

The mass contribution from such stars is small.
Assuming the $M/L_J$ ratio of an evolved population ($\simgt\,$1\,Gyr)
to be 1.1 in solar units (see, e.g., \citet{moriondo}), 
and that of intermediate-age (100\,--\,500\,Myr) 
and young stars ($\simlt\,$50\,Myr) to be respectively 0.5 and 0.1, 
we calculate the fraction of the total stellar mass that can be in the 
form of old stars in the NICMOS CMD. To
conservatively account for incompleteness, 
we assign 4\% of the total stellar light 
to an old stellar population of age $\simgt\,$1\,Gyr, 
twice as large as the 2\% fraction observed in the CMD.
Using the fractions in Table \ref{table:par}, we then adjust the light of the 
intermediate-age population (by subtracting 0.04 from $r_*(J)$), and
calculate for each region
the fraction of the stellar mass that could reside in old stars.
We do not consider those regions
where gas contributes more than 70\% to the $J$ light. 
We find that the total stellar mass in \izw\ in old stars could be at 
most 22\%; this high fraction is found in the SE1 and Inter. regions 
where half or more of the light comes from young stars.
If we consider an extreme case (not supported by our data), with no gas and
70\% of the light in young stars, the 4\% $J$ light fraction in old stars
would correspond to 18\% of the stellar mass.
Thus, even in the least favorable scenario, the stellar mass fraction in old stars does not exceed 20\%.

With deep ground-based CCD and NIR images of \izw,
\citet{ko00} find evidence for a $B-J$ color that increases
with radius, reaching $B-J\,\simeq\,1.5$ at $R$\,=\,10\arcsec.
They attribute the red $B-J$ color to a stellar population with 
an age of more than 1\,Gyr.
While we confirm the color trend (see Figs. \ref{fig:pro} and \ref{fig:cut}),
we disagree with the interpretation.
Our five-color fits are more consistent with a substantial (60-80\%) contribution
from ionized gas in the extended regions.
As mentioned above, this is because we have included the $V-I$ color in our fits;
without the $V-I$ color, the red $B-J$ color mimics stars of age $\simgt\,$1\,Gyr.
A recent photometric analysis including optical colors by \citet{papa02} also 
finds the extended regions surrounding \izw\ to be highly contaminated by gas.
Taking the gas emission into account, and with an independent analysis,
they also derive an upper limit of $\sim$\,500\,Myr for the age of \izw. 

We next compare our findings to those of \citet{aloisi} who 
found, from their CMD study of 
\hst/WFPC2 images, evidence for two bursts.
The first, older episode, ``started any time between 1 and 0.2\,Gyr ago'',
while the second, more recent, one has an age between 15 and 20\,Myr. 
Given that they deliberately excluded the brightest H{\sc ii} regions and unresolved
star clusters from their analysis,
their age estimate of the of the young burst and ours agree rather well.
However, the older age estimate depends on the distance adopted for \izw.
The difference between our older age estimate of $\sim$\,500\,Myr and their upper limit of 1\,Gyr
could be reduced
if the distance to \izw\ were 12.6\,Mpc \citep{ostlin} instead of 10\,Mpc as they assumed.
\citet{aloisi}'s derived age of 200\,Myr for the C component
(with large error bars as the data are of limited sensitivity) 
is in good agreement with our result.

Recent work on the chemical abundances and the star-formation history in \izw\
has also suggested that there were two bursts of star formation, one which occurred 
300\,Myr ago, and a young one with age of 4-7\,Myr \citep{recchi}.
Our age estimates of the two stellar populations in \izw\ are in very good agreement
with these ages.
Considering the two radically different approaches, this is a strong confirmation of
the young age ($\simlt\,$500\,Myr) of the oldest major stellar population 
in \izw.

\section{Conclusions \label{concl}}

We have presented deep UKIRT $JHK$ images of the Blue Compact Dwarf 
\izw\ and analyzed these 
images in conjunction with archival \hst/WFPC2 optical images.
By dividing the main body into eight different regions, and fitting five
broadband colors with evolutionary synthesis models to these and the
C component, we are able
to set an upper limit to the age of the major component of the 
old stellar population in \izw. 
We find the following:

\begin{enumerate}
\item
The oldest major  
stellar population in \izw, contributing $\geq$ 78\% of its mass, 
is at most 500\,Myr old.
Within the uncertainties of our data, by
fitting the optical, NIR, and optical-NIR colors with evolutionary synthesis 
models of various star formation histories we are able to set an upper limit of 
$\sim$ 500 Myr for the age of the oldest stars in this major component. 
However, our data are insensitive to as much as a 22\% contribution in
mass (4\% in $J$ light) from stars older than 500\,Myr in \izw.
\item
The young stellar populations in the main body of \izw\ and the C component
range from 3 to 15\,Myr in age. 
The largest fractions of older stars appear to be found in the C component
and in the SE star cluster;
in the remaining regions, young stars and gas contribute half or more 
of the observed light.
\item
The observed broadband colors in most regions of \izw\  
are significantly affected by ionized gas emission and dust extinction.
Red $H-K$, $B-H$, and $V-K$ are not reliable indicators of old stellar populations with
red colors because ionized gas emission is also red.
$V-I$, on the other hand, reliably separates stars from gas because 
the $V-I$ color of stars is red ($\geq$ 0.4) while that of gas is blue 
($\sim$ --0.4).
\end{enumerate}

Thus the question of the 
age of \izw\ is not settled by our deep NIR observations. While 
we have shown that the oldest major stellar population in \izw\ is at most 
500\,Myr old, we cannot exclude the possibility that as much as 22\% 
of the mass of \izw\ is contributed by stars older than 500\,Myr. The advent 
of the Advanced Survey Camera (ACS) 
on \hst\ will allow to settle definitively the debate.    
Deep ACS images with higher spatial resolution and 
sensitivity than WFPC2 will enable detection of or at least constrain the red giant 
branch (RGB) stellar population in \izw. If RGB stars are not detected, then 
we can set an upper limit for the age of \izw\ to be less than $\sim$ 1 Gyr.
If they are detected, \izw\ is not young, and the RGB tip can be used to 
derive its distance.   

\acknowledgments
We thank Andy Adamson and Sandy Leggett of UKIRT for their
competent observing, and the UKIRT Time Allocation Panel for
their generous allocation of observing time.
Polis Papaderos graciously communicated his results to us in advance 
of publication. 
L.K.H.'s research was partially funded by ASI Grant ASI I/R/35/00.
T.X.T. and Y.I.I. thank the partial financial support of 
National Science Foundation grant AST-02-05785. T.X.T is also 
grateful for the support of \hst\ grant GO-08769.01-A.

\clearpage

\onecolumn

\clearpage

\singlespace

\begin{deluxetable}{lrrrrrrr}
\tabletypesize{\scriptsize}
\tablecaption{NIR aperture photometry of \izw \label{table:phot}}
\tablewidth{0pt}
\tablehead{ & \multicolumn{1}{c}{Aperture} \\
\colhead{Region} & \colhead{$\phi$ (\arcsec)} &
\colhead{{\it F555W}\tablenotemark{a}} & \colhead{$J$} & \colhead{$H$} & \colhead{$K$} &
\colhead{$V-I$\tablenotemark{a}} & \colhead{$B-H$\tablenotemark{a}} 
}
\startdata
NW2
	&  2 & 21.178 ( 0.003) &  21.124 ( 0.405) &  21.035 ( 0.481) &  20.258 ( 0.420) &   -0.400 &  0.743\\
      &  3 & 20.130 ( 0.002) &  20.055 ( 0.248) &  19.776 ( 0.269) &  19.270 ( 0.267) &   -0.368 &  0.928\\
      &  4 & 19.252 ( 0.001) &  19.226 ( 0.169) &  18.949 ( 0.183) &  18.456 ( 0.183) &   -0.316 &  0.821\\
NW1
	&  2 & 19.316 ( 0.001) &  19.315 ( 0.176) &  18.970 ( 0.186) &  18.722 ( 0.208) &   -0.128 &  0.635\\
      &  3 & 18.382 ( 0.001) &  18.343 ( 0.113) &  17.969 ( 0.117) &  17.795 ( 0.135) &   -0.026 &  0.645\\
      &  4 & 17.743 ( 0.001) &  17.662 ( 0.083) &  17.317 ( 0.087) &  17.143 ( 0.100) &   -0.019 &  0.591\\
NW
	&  2 & 17.709 ( 0.001) &  17.681 ( 0.083) &  17.413 ( 0.091) &  17.318 ( 0.109) &    0.004 &  0.298\\
      &  3 & 17.300 ( 0.001) &  17.179 ( 0.066) &  16.906 ( 0.072) &  16.778 ( 0.085) &   -0.011 &  0.428\\
      &  4 & 17.049 ( 0.001) &  16.894 ( 0.058) &  16.618 ( 0.063) &  16.457 ( 0.073) &   -0.033 &  0.506\\
      &  6 & 16.772 ( 0.001) &  16.568 ( 0.050) &  16.303 ( 0.055) &  16.054 ( 0.061) &   -0.067 &  0.603\\
      &  8 & 16.611 ( 0.001) &  16.357 ( 0.045) &  16.080 ( 0.049) &  15.824 ( 0.055) &   -0.058 &  0.696\\
      & 10 & 16.481 ( 0.001) &  16.190 ( 0.042) &  15.914 ( 0.045) &  15.643 ( 0.050) &   -0.061 &  0.750\\
      & 16 & 16.244 ( 0.001) &  15.872 ( 0.036) &  15.578 ( 0.039) &  15.315 ( 0.043) &   -0.067 &  0.894\\
      & 20 & 16.218 ( 0.001) &  15.815 ( 0.035) &  15.506 ( 0.038) &  15.271 ( 0.042) &   -0.078 &  0.953\\
Int.
	&  2 & 19.125 ( 0.001) &  18.955 ( 0.150) &  18.720 ( 0.165) &  18.356 ( 0.175) &   -0.154 &  0.592\\
      &  3 & 18.238 ( 0.001) &  18.007 ( 0.096) &  17.766 ( 0.107) &  17.445 ( 0.115) &   -0.098 &  0.592\\
      &  4 & 17.547 ( 0.001) &  17.331 ( 0.071) &  17.074 ( 0.078) &  16.812 ( 0.086) &   -0.055 &  0.584\\
SE
	&  2 & 19.352 ( 0.001) &  18.723 ( 0.134) &  18.383 ( 0.142) &  18.223 ( 0.165) &    0.170 &  1.174\\
      &  3 & 18.615 ( 0.001) &  18.074 ( 0.099) &  17.750 ( 0.106) &  17.532 ( 0.120) &    0.073 &  1.086\\
      &  4 & 18.041 ( 0.001) &  17.584 ( 0.079) &  17.301 ( 0.086) &  17.042 ( 0.096) &    0.012 &  0.975\\
      &  6 & 17.373 ( 0.001) &  16.910 ( 0.058) &  16.645 ( 0.063) &  16.352 ( 0.069) &   -0.035 &  0.972\\
      &  8 & 16.808 ( 0.001) &  16.386 ( 0.046) &  16.110 ( 0.050) &  15.850 ( 0.055) &   -0.030 &  0.864\\
      & 10 & 16.465 ( 0.001) &  16.077 ( 0.039) &  15.779 ( 0.042) &  15.550 ( 0.048) &   -0.008 &  0.842\\
      & 16 & 16.246 ( 0.001) &  15.854 ( 0.036) &  15.547 ( 0.038) &  15.313 ( 0.043) &   -0.047 &  0.910\\
      & 20 & 16.221 ( 0.001) &  15.815 ( 0.035) &  15.509 ( 0.038) &  15.275 ( 0.042) &   -0.061 &  0.954\\
SE1
	&  2 & 19.041 ( 0.001) &  18.837 ( 0.141) &  18.665 ( 0.161) &  18.359 ( 0.175) &   -0.160 &  0.610\\
      &  3 & 18.470 ( 0.001) &  18.163 ( 0.103) &  17.962 ( 0.117) &  17.632 ( 0.126) &   -0.108 &  0.718\\
      &  4 & 18.137 ( 0.001) &  17.714 ( 0.084) &  17.459 ( 0.093) &  17.193 ( 0.103) &   -0.053 &  0.901\\
SE2
	&  2 & 21.439 ( 0.004) &  20.409 ( 0.291) &  20.141 ( 0.318) &  19.763 ( 0.334) &    0.145 &  1.183\\
      &  3 & 20.118 ( 0.002) &  19.323 ( 0.177) &  19.020 ( 0.190) &  18.788 ( 0.213) &    0.121 &  1.149\\
      &  4 & 19.358 ( 0.001) &  18.645 ( 0.129) &  18.401 ( 0.143) &  18.137 ( 0.158) &    0.077 &  1.082\\
\cutinhead{C component}
C(SE)
	&  2 &&  20.944 ( 0.062) &  20.792 ( 0.240) &  20.643 ( 0.334) &    &      \\
      &  3 &&  20.475 ( 0.060) &  19.979 ( 0.169) &  20.046 ( 0.288) &    &      \\
      &  4 &&  20.072 ( 0.056) &  19.564 ( 0.154) &  19.365 ( 0.205) &    &      \\
C(NW)
	&  2 &&  21.414 ( 0.096) &  20.830 ( 0.248) &  20.745 ( 0.367) &    &      \\
      &  3 &&  20.729 ( 0.076) &  20.059 ( 0.183) &  20.232 ( 0.342) &    &      \\
      &  4 &&  20.209 ( 0.063) &  19.560 ( 0.154) &  19.633 ( 0.263) &    &      \\
C(Int.=Tot.)
	 &  2 &&  21.216 ( 0.080) &  20.888 ( 0.261) &  20.381 ( 0.263) &  0.328 &  0.507 \\
      &  3 &&  20.363 ( 0.055) &  19.872 ( 0.153) &  19.705 ( 0.211) \\
      &  4 &&  19.909 ( 0.048) &  19.457 ( 0.140) &  19.253 ( 0.186) \\
      &  6 &&  19.433 ( 0.046) &  18.999 ( 0.137) &  18.820 ( 0.186) \\
      &  8 &&  19.145 ( 0.047) &  18.568 ( 0.123) &  18.470 ( 0.239) \\
      & 10 &&  18.831 ( 0.044) &  18.313 ( 0.121) &  18.270 ( 0.304) \\
      & 16 & 19.20\tablenotemark{b} (0.100) &  18.311 ( 0.044) &  17.888 ( 0.132) &  17.971 ( 0.257) 
&  0.12\tablenotemark{b} &   1.30\\
\enddata
\tablenotetext{a}{The values of {\it F555W}, $V-I$, and $B-H$ have {\it not} been transformed
to ground-based photometry in this table; for reasons of brevity, 
$B$ has been used to denote {\it F439W}, $V$ for {\it F555W}, and $I$ for {\it F814W}.
Colors here have not been corrected for Galactic extinction.}
\tablenotetext{b}{From \citet{papa02}.}
\end{deluxetable}

\clearpage


\begin{deluxetable}{llcrr}
\tablenum{2}
\tablecolumns{5}
\tablecaption{Model fits\tablenotemark{a} ~to colors of \izw \ C Component\label{table:fitc}}
\tablewidth{0pt}
\tablehead{
\colhead{Old Age\tablenotemark{b}} & 
\colhead{Young Age\tablenotemark{c}} &
\colhead{$r_*(J)$} & \colhead{$\Sigma\,\chi^2$} & \colhead{RMS (mag)\tablenotemark{d}} \\
\cline{1-5} \\
\multicolumn{5}{c}{Young Instantaneous Burst (IB)\tablenotemark{c}} } 
\startdata
100-200Myr  & 15Myr & 0.6 & 16.7 & 0.21 \\
	    & 20Myr & 0.5 & 17.3 & 0.19 \\
100-500Myr  & 15Myr & 0.6 & 17.4 & 0.22 \\
10-200Myr   & 15Myr & 0.9 & 17.5 & 0.18 \\
            & 20Myr & 0.8 & 17.7 & 0.20 \\
100-500Myr  & 20Myr & 0.5 & 17.8 & 0.19 \\
10-500Myr   & 15Myr & 0.7 & 17.8 & 0.22 \\
3-500Myr    & 15Myr & 0.8 & 17.9 & 0.19 \\
10-500Myr   & 20Myr & 0.6 & 18.1 & 0.20 \\
3-500Myr    & 20Myr & 0.7 & 18.2 & 0.19 \\
\cutinhead{Young Continuous Burst (CB)\tablenotemark{c}}
100-200Myr  & 10-20Myr & 0.6 & 17.0 & 0.23 \\
            & 10-20Myr & 0.7 & 17.4 & 0.16 \\
10-200Myr   & 10-20Myr & 0.9 & 17.4 & 0.19 \\
3-500Myr    & 10-20Myr & 0.8 & 17.6 & 0.20 \\
100-200Myr  & 15-20Myr & 0.5 & 17.6 & 0.22 \\
100-500Myr  & 10-20Myr & 0.6 & 17.7 & 0.23 \\
10-200Myr   & 15-20Myr & 0.8 & 17.7 & 0.21 \\
100-500Myr  & 10-20Myr & 0.7 & 17.8 & 0.16 \\
100-200Myr  & 15-20Myr & 0.6 & 17.9 & 0.17 \\
10-500Myr   & 10-20Myr & 0.7 & 17.9 & 0.23 \\
\enddata
\tablenotetext{a}{Fitted young-population colors ($V-I$, $B-H$, $V-K$, $J-H$, $H-K$) of C component
are stars only: gas emission and extinction are constrained to be zero.}
\tablenotetext{b}{Evolved population colors assume a continuous star-formation scenario. }
\tablenotetext{c}{Scenarios refer to the young population.}
\tablenotetext{d}{Root-mean-square residual (in mag) averaged over all five colors.}
\end{deluxetable}

\clearpage


\begin{deluxetable}{llllllllllr}
\tablenum{3}
\tabletypesize{\scriptsize}
\tablecaption{Ages of Model Fits\tablenotemark{a} ~to Colors of \izw \label{table:fit}}
\tablewidth{0pt}
\tablehead{
\colhead{Old Age\tablenotemark{b}} &
\colhead{NW2} & \colhead{NW1} & \colhead{NW} & \colhead{Inter.} & 
\colhead{SE} & \colhead{SE1} & \colhead{SE2} & 
\colhead{Ext.} & \colhead{C Comp.} & \colhead{$\Sigma\,\chi^2$\tablenotemark{c}}
}
\startdata
100-200Myr (IB)\tablenotemark{d} 
                & 3Myr    & 3Myr    & 3Myr   & 3Myr   & 10Myr  & 3Myr   & 10Myr  & 10Myr   & 15Myr   &  64.6 \\
	        &  12.4 &  13.0 &   1.9 &   2.1 &   3.0 &   5.6 &   2.9 &   6.8 &  16.7  \\
100-500Myr (IB) & 3Myr    & 3Myr    & 3Myr   & 3Myr   & 3Myr   & 3Myr   & 10Myr  & 10Myr   & 15Myr   &  67.2 \\
		&  12.4 &  14.2 &   2.1 &   2.4 &   2.8 &   5.7 &   3.1 &   7.0 &  17.4  \\
100-200Myr (CB) & 3Myr    & 20Myr   & 3Myr   & 20Myr  & 3-20Myr & 15Myr  & 3-20Myr & 100Myr  & 10-20Myr &  69.9 \\
		&  12.0 &  16.6 &   1.9 &   2.9 &   2.6 &   7.0 &   3.0 &   6.8 &  17.0  \\
10-500Myr (IB)  & 3Myr    & 3Myr    & 3Myr   & 3Myr   & 10Myr  & 3Myr   & 10Myr  & 10Myr   & 15Myr   &  70.4 \\
	        &  12.4 &  16.6 &   2.0 &   2.5 &   2.8 &   5.7 &   3.0 &   7.6 &  17.8  \\
10-200Myr (IB) & 3Myr    & 3Myr    & 10Myr  & 3Myr   & 3Myr   & 3Myr   & 10Myr  & 10Myr   & 15Myr   &  72.5 \\
                 &  12.4 &  16.6 &   2.8 &   2.8 &   2.9 &   5.8 &   3.6 &   8.0 &  17.5  \\
100-500Myr (CB) & 3Myr    & 20Myr   & 3Myr   & 20Myr  & 3-20Myr & 15Myr  & 3-20Myr & 100Myr  & 10-20Myr &  72.7 \\
                &  12.0 &  17.6 &   2.2 &   3.2 &   2.6 &   7.2 &   3.4 &   7.0 &  17.7  \\
3-500Myr (IB) & 3Myr    & 3Myr    & 3Myr   & 3Myr   & 10Myr  & 3Myr   & 10Myr  & 10Myr   & 15Myr   &  73.0 \\
                 &  12.4 &  17.3 &   2.0 &   2.8 &   3.1 &   6.0 &   3.6 &   7.9 &  17.9  \\
3Myr-1Gyr   (IB) & 3Myr    & 3Myr    & 3Myr   & 3Myr   & 10Myr  & 3Myr   & 10Myr  & 10Myr   & 15Myr   &  73.1 \\
                 &  12.4 &  17.5 &   2.2 &   2.8 &   2.9 &   5.9 &   3.2 &   7.7 &  18.4  \\
100Myr-1Gyr (IB) & 3Myr    & 3Myr    & 3Myr   & 3Myr   & 3Myr   & 3Myr   & 10Myr  & 10Myr   & 15Myr   &  73.9 \\
                 &  12.4 &  16.5 &   2.7 &   2.9 &   3.1 &   6.1 &   3.8 &   7.4 &  18.8  \\
100Myr-2Gyr (IB) & 3Myr    & 3Myr    & 10Myr  & 3Myr   & 10Myr  & 3Myr   & 10Myr  & 10Myr   & 15Myr   &  74.5 \\
                 &  12.4 &  16.6 &   2.6 &   3.1 &   3.1 &   6.4 &   3.9 &   7.3 &  19.1  \\
10-500Myr (CB) & 3Myr    & 15Myr   & 3Myr   & 20Myr  & 3-15Myr & 15Myr  & 3-20Myr & 100Myr  & 10-20Myr &  75.0 \\
                 &  12.0 &  18.5 &   2.2 &   3.6 &   2.6 &   7.2 &   3.4 &   7.6 &  17.9  \\
0-1Gyr      (IB) & 3Myr    & 3Myr    & 3Myr   & 3Myr   & 10Myr  & 3Myr   & 10Myr  & 10Myr   & 15Myr   &  75.1 \\
                 &  12.4 &  18.1 &   2.4 &   3.0 &   3.0 &   6.0 &   3.4 &   7.8 &  18.8  \\
10-200Myr (CB) & 3Myr    & 15Myr   & 3Myr   & 3Myr-10Myr & 20Myr  & 20Myr  & 100Myr & 100Myr  & 10Myr-20Myr &  75.9 \\
                 &  12.0 &  18.7 &   2.3 &   4.0 &   2.9 &   7.4 &   3.3 &   8.0 &  17.4  \\
3-500Myr (CB) & 3Myr    & 15Myr   & 3Myr   & 20Myr  & 3-20Myr & 15Myr  & 3-20Myr & 100Myr  & 10-20Myr &  76.3 \\
                 &  12.0 &  18.8 &   2.4 &   3.8 &   3.0 &   7.4 &   3.4 &   7.9 &  17.6  \\
100Myr-2Gyr (CB) & 3Myr    & 20Myr   & 3Myr   & 20Myr  & 20Myr  & 15Myr  & 3-20Myr & 100Myr  & 10-20Myr &  77.8 \\
                 &  12.0 &  19.7 &   2.1 &   3.8 &   2.8 &   7.8 &   3.8 &   7.3 &  18.5  \\
3Myr-1Gyr   (CB) & 3Myr    & 15Myr   & 3Myr   & 20Myr  & 3-10Myr & 15Myr  & 3-20Myr & 100Myr  & 10-20Myr &  77.9 \\
                 &  12.0 &  19.5 &   2.3 &   3.8 &   3.0 &   7.5 &   3.9 &   7.7 &  18.3  \\
100Myr-1Gyr (CB) & 3Myr    & 20Myr   & 3Myr   & 20Myr  & 3-20Myr & 15Myr  & 3-20Myr & 100Myr  & 10-20Myr &  78.7 \\
                 &  12.0 &  19.4 &   2.4 &   3.7 &   3.2 &   7.7 &   4.1 &   7.4 &  18.8  \\
0-1Gyr      (CB) & 3Myr    & 15Myr   & 3Myr   & 20Myr  & 20Myr  & 15Myr  & 3-20Myr & 100Myr  & 10-20Myr &  79.3 \\
                 &  12.0 &  19.9 &   2.8 &   3.9 &   2.8 &   7.5 &   3.9 &   7.8 &  18.8  \\
\enddata
\tablenotetext{a}{Fitted colors ($V-I$, $B-H$, $V-K$, $J-H$, $H-K$) of young population
include gas emission as described in text. 
For each region, the young age is reported in the first line, and
$\chi^2$ (summed over all five colors) in the second.}
\tablenotetext{b}{All evolved population colors assume a continuous star-formation scenario. }
\tablenotetext{c}{$\chi^2$ (over all five colors) summed over all regions.
There may be small discrepancies with respect to the sum of the values in the second line,
because of numerical round-off in the latter.}
\tablenotetext{d}{The scenarios refer to the young population.} 
\end{deluxetable}

\clearpage


\begin{deluxetable}{llllllrrrrr}
\tablenum{4}
\tablecolumns{11}
\tabletypesize{\scriptsize}
\tablecaption{Parameters and Colors of Best-Fit Models\tablenotemark{a} ~of \izw \label{table:par}}
\tablewidth{0pt}
\tablehead{ 
\colhead{Region} & \colhead{Young Age\tablenotemark{b}} & 
\colhead{$A_V$} & \colhead{$r_*(J)$\tablenotemark{c}} & \colhead{$r_{\rm gas}(J)$} & \colhead{RMS\tablenotemark{d}} &
\colhead{$J-H$} & \colhead{$H-K$} & \colhead{$V-I$} & \colhead{$V-K$} & \colhead{$B-H$} 
}
\startdata
NW2   & 3Myr  & 0.0 & 0.0 & 0.8 &  0.23 &  0.01 ( 1.71) &  0.60 (-0.52) & -0.26 (-2.63) &  1.15 (-1.36) &  0.60 ( 0.70) \\
      & & & & & 0.21 &   0.27 ( 0.15) &  0.49 ( 0.20) & -0.39 ( 0.05) &  0.78 ( 0.27) &  0.79 ( 0.27) \\
NW1   & 3Myr  & 0.0 & 0.4 & 0.3 &  0.14 &  0.21 ( 2.58) &  0.24 ( 0.01) & -0.05 (-2.18) &  0.67 (-0.79) &  0.32 ( 1.00) \\
      & & & & & 0.13 &   0.34 ( 0.05) &  0.24 ( 0.07) & -0.16 ( 0.05) &  0.50 ( 0.21) &  0.51 ( 0.19) \\
NW    & 3Myr  & 0.1 & 0.5 & 0.0 &  0.05 &  0.25 ( 0.19) &  0.06 ( 0.56) &  0.00 (-0.98) &  0.38 (-0.75) &  0.16 ( 0.17) \\
      & & & & & 0.07 &   0.26 ( 0.05) &  0.09 ( 0.05) & -0.04 ( 0.05) &  0.29 ( 0.11) &  0.18 ( 0.09) \\
Inter. & 3Myr & 0.0 & 0.3 & 0.3 &  0.06 &  0.17 ( 0.48) &  0.33 ( 0.17) & -0.13 (-1.20) &  0.77 (-0.47) &  0.39 ( 0.45) \\
      & & & & & 0.12 &   0.19 ( 0.05) &  0.34 ( 0.05) & -0.19 ( 0.05) &  0.68 ( 0.18) &  0.47 ( 0.17) \\
SE    & 10Myr & 0.1 & 0.6 & 0.2 &  0.08 &  0.33 ( 0.00) &  0.21 (-1.08) &  0.14 (-0.48) &  1.08 (-0.36) &  0.88 ( 1.22) \\
      & & & & & 0.10 &   0.33 ( 0.05) &  0.15 ( 0.05) &  0.12 ( 0.05) &  1.02 ( 0.16) &  1.05 ( 0.14) \\
SE1   & 3Myr  & 0.0 & 0.3 & 0.4 &  0.11 &  0.16 (-0.05) &  0.31 (-0.40) & -0.10 (-1.92) &  0.72 (-0.75) &  0.31 ( 1.10) \\
      & & & & & 0.11 &   0.16 ( 0.05) &  0.29 ( 0.05) & -0.19 ( 0.05) &  0.59 ( 0.17) &  0.49 ( 0.16) \\
SE2   & 10Myr & 0.0 & 0.4 & 0.5 &  0.08 &  0.24 ( 0.88) &  0.36 (-1.22) &  0.10 (-0.52) &  1.27 (-0.18) &  0.93 ( 0.55) \\
      & & & & & 0.14 &   0.30 ( 0.07) &  0.22 ( 0.11) &  0.08 ( 0.05) &  1.23 ( 0.21) &  1.03 ( 0.19) \\
Exten. & 10Myr & 0.2 & 0.4 & 0.6 &  0.31 &  0.26 ( 0.96) &  0.42 (-0.50) &  0.11 (-2.16) &  1.59 ( 0.26) &  1.39 ( 0.95) \\
      & & & & & 0.29 &   0.38 ( 0.13) &  0.32 ( 0.20) & -0.43 ( 0.25) &  1.69 ( 0.38) &  1.77 ( 0.40) \\
C      & 15Myr & 0.0 & 0.6 & 0.0 &  0.21 &  0.32 ( 0.70) &  0.09 (-0.67) &  0.22 (-2.89) &  0.90 ( 0.92) &  0.85 ( 2.56) \\
      & & & & & 0.19 &   0.41 ( 0.13) & -0.09 ( 0.27) &  0.08 ( 0.05) &  1.13 ( 0.26) &  1.18 ( 0.13) \\
\enddata
\tablenotetext{a}{For each region, best-fit colors ($V-I$, $B-H$, $V-K$, $J-H$, $H-K$) are given on first
line (with parentheses giving residual weighted by photometric uncertainties), 
and observed colors on second line (with photometric uncertainties in parentheses).
All observed colors have been corrected for Galactic extinction as described in the text.}
\tablenotetext{b}{The model colors of the young population include gas emission as described in text,
and an IB since it gives the best fit.}
\tablenotetext{c}{Model colors of the evolved population assume a CB of age 100--200Myr.}
\tablenotetext{d}{Root-mean-square residual (in mag) averaged over all five colors.}
\end{deluxetable}

\clearpage

\end{document}